\documentclass[ ]{aa} 
\usepackage{graphicx}          
\usepackage{natbib}

\newcommand{\bp}{$\beta$~Pic}

\begin{document} 
\title{Survival of icy grains in debris discs. }
\subtitle{The role of photosputtering} 
 \offprints{A.\ Grigorieva} \mail{anja@astro.su.se}
 \date{Received / accepted} \titlerunning{Survival of icy grains  
 in debris discs}                       
\author{Anna Grigorieva\inst{1}
  \and Ph. Th\'ebault\inst{1,3}   
 \and P. Artymowicz\inst{2}
 \and A. Brandeker\inst{1}}
\institute{Stockholm Observatory, SCFAB, SE-10691 Stockholm, Sweden
\and 
University of Toronto at Scarborough, 1265 Military Trail, 
Toronto, Ontario, M1C~1A4,  Canada
\and
Observatoire de Paris, Section de Meudon,
F-92195 Meudon Principal Cedex, France}
 \date{Received / accepted} \titlerunning{Collisional avalanches in
 dusty discs}                       
\authorrunning{A.\ Grigorieva et al.}
\abstract
{}
{We put theoretical constraints on the presence
and survival of icy grains in debris discs. Particular attention is paid
to UV sputtering of water ice, which has so far not been 
studied in detail in this context.}
{We present a photosputtering model based on
available experimental and theoretical studies.
We quantitatively estimate the erosion rate of icy and
ice-silicate grains, under the influence of both sublimation and
photosputtering, as a function of grain size, composition and
distance from the star.
The effect of erosion on the grain's location is investigated
through numerical simulations coupling the grain size to
its dynamical evolution. }
{Our model predicts that photodesorption efficiently destroy 
 ice in optically thin discs, even far beyond
the sublimation snow line. For the reference case of 
$\beta$~Pictoris, we find that only $\ga 5\,$\,mm grains can keep
their icy component for the age of the system in the 50--150\,AU region.
When taking into account the collisional reprocessing 
of grains, we show that the water ice survival on grains 
improves (grains down to $\simeq 20\,\mu$m might be partially icy).
However, estimates of the amount of gas photosputtering
would produce on such a hypothetical population of big icy grains
lead to values for the O\,I column density that strongly exceed
observational constraints for $\beta$~Pic, thus 
ruling out the presence of a significant amount of icy grains
in this system.
Erosion rates and icy grains survival timescales are also
given for a set of 11 other debris disc systems. We show that,
with the possible exception of M stars, photosputtering cannot
be neglected in calculations of icy grain lifetimes. 
 }
{}

\maketitle
 
\section{Introduction } \label{sec:introduction}

 Water ice has been unambiguously detected 
towards embedded YSOs/protostars 
\citep[][ and references therein]{Gibb04} and around 
 Herbig~AeBe and  T~Tauri stars
\citep[e.g.,][]{Creech02, Meeus01, Gibb04}.
For the more evolved debris disc systems, however, direct
observational clues for the
presence of icy grains seem far less conclusive.
The obvious exception is of course our own Solar System where, in addition to
appearing on satellite and planetary surfaces, dust-sized
ice particles have been detected in the comae of comets and in planetary rings.
For extrasolar systems, the presence of ice
appears very likely, given the fact that
many debris discs extend to cold regions very far from their central stars,
beyond the expected sublimation limit.
However, no unambiguous ``smoking gun''  (presumably water gun?)
signature of their presence is available yet.
Water ice is usually inferred through indirect arguments like
best fits of the Spectral Energy Distribution (SED).
The most conclusive of these indirect detections
are probably the ones around the K stars 
$\epsilon$\,Eridani (0.5--1\,Gyr) \citep{Li03Eri} and
HD\,69830 \citep{Lisse07}, for which the presence of water ice 
appears as a robust conclusion.

For discs around more luminous stars, definitive conclusions are more
difficult to reach and appear more model dependent.
\citet[][]{Chen06} analyzed spectral energy distributions (SEDs)
around A, B and F~types stars with detected infrared excesses 
and find that, when the excesses are
reproduced by a single temperature blackbody  emission,
the peak in the number of systems
with a given grain temperature occurs at 110\,--\,120\,K. 
This was interpreted by the authors as
a possible indication of the icy nature of the grains, since
sublimation timescales for warmer grains are much shorter than
the estimated system ages. The infrared (IR) emission detected in
the HR\,4796 system  (an A-type star) was also reproduced 
by 110\,K blackbody grains  \citep[][]{Jura98}.
Absence of warmer grains was again interpreted as a result of
sublimation of ice in the inner part of the disc
(however, as another possible explanation, a sweeper  companion was
also mentioned).
\citet[][]{Golimowski06} invoked ice to explain the 
observed color change beyond 120\,AU in the archetypal \bp~disc 
(again around an A-type star). Their idea is that 
beyond the sublimation zone the average grain size is 
bigger, since the grains can preserve icy mantles.

However, even if the observational evidence is not 100\%
conclusive, there are numerous theoretical arguments supporting
the possible presence of ice in debris discs. As already mentioned,
the most obvious one is that debris discs often extend to regions
beyond the sublimation snowline, i.e.
thought to be cold enough for water vapor to condense on dust grains, the
exact location of the sublimation zone depending on the thermal balance at
the grain surfaces. Furthermore, since ices have been detected around YSOs, it
should be logical to expect it to still be there at later times, 
provided no new ice removing mechanism appeared along the way. 
\citet{Li98} also argued that if grains are produced from
cometary evaporation then, since comet dust may consist 
of primitive interstellar
dust on which the presence of an icy mantle is a well established fact, we
should expect evaporated volatiles to recondense  
in the outer regions of debris discs
(beyond $\simeq$100\,AU for the \bp~case considered by these authors), or
the ice mantles to be preserved for dust particles produced by 
shattering or collisions.

In this paper, we revisit the theoretical constraints 
on the  presence  of ice in debris discs. 
Our main point is that sublimation is not the only process 
which leads to ice destruction. We focus here on
UV sputtering (a.k.a. photosputtering 
or photodesorption) of circumstellar ice particles,
a  phenomenon well known in the context of interstellar grains.
In the solar system, the ice 
destruction rate due to UV radiation was estimated to  
 be $\sim 0.4$\,cm\,Myr$^{-1}$ at the distance of Saturn 
 \citep{Harrison67,Carlson80}. 
In his early pioneering work, \citet{Art94, Art96} predicted this rate
to be orders of magnitude higher around A-type stars,
which have a much stronger UV flux.
In our paper, we provide a detailed model of
the UV sputtering mechanism (Sect.~\ref{sec:sp}). It 
extends previous calculations (which were
not described in detail by Artymowicz), to include  
new laboratory and theoretical results that have become available in the last
decade, as well a range of different host stars.
 In addressing the issue of ice survival in circumstellar discs, we 
also include sublimation (Sect.~\ref{sec:sub}) in our model.
In Sect.~\ref{sec:dynamics} we consider how grain dynamics is 
affected by erosion of the grain surface. 
Section~\ref{sec:bp} deals with application of our study to 
the \bp~system.   
Sputtering rates for 11 other systems with detected debris discs 
can be found in Sect.~\ref{sec:other}.
Summary and Conclusions are presented in Sect.~\ref{sec:summary}.

\section{Model description} 
\label{sec:model}

\begin{figure}[t]
\includegraphics[width=\columnwidth,clip]{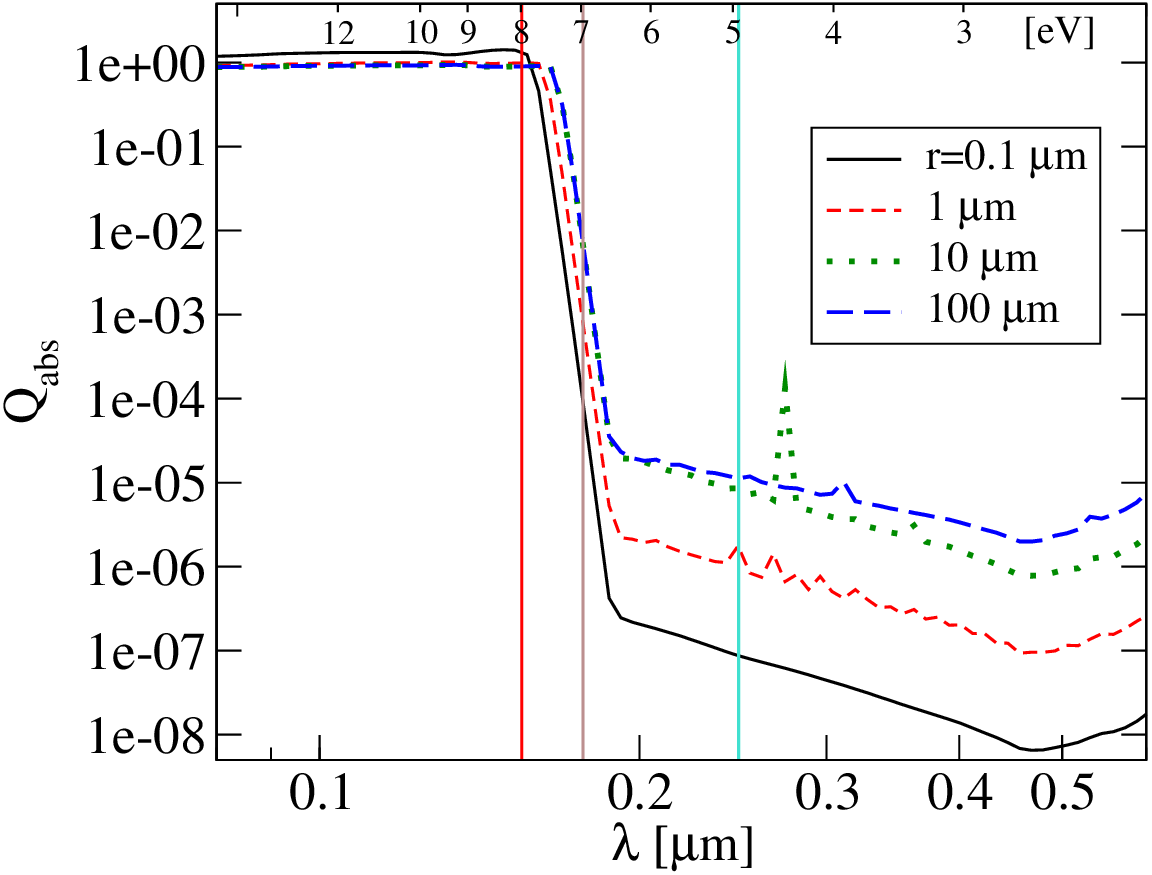}
\caption{Absorption coefficients for water ice grains of different sizes.
(Mie theory code of Artymowicz 1988.) }
\label{fig:H2O_Qabs}
\end{figure} 

In order to address the issue of ice survival in debris discs around 
stars we consider, in that order:
(i) sublimation, (ii) photosputtering, and (iii) related grain dynamics.
We compare the erosion timescales with the system age 
as well as the collisional lifetime of grains, the latter being importantly 
a source of newly generated dust grains.

\subsection{Sublimation {\label{sec:sub}}}

Sublimation is a well known process.
The mass loss  rate $\Phi$ is proportional to the 
difference between the phase-equilibrium vapor pressure $p_\mathrm{sat}$ and 
the actual water vapor pressure  $p$ above the icy surface, 
i.e. $\Phi \propto (p_\mathrm{sat} - p)$.\footnote{
Condensation occurs when  $p > p_\mathrm{sat}$.}
In our calculations,   
we neglect  the possible presence of water molecules in the ambient gas,
assuming $p=0$ 
(an assumption which is justified in a debris disc).
In this case, the mass sublimation rate can be written as \citep{Lamy74}
\begin{eqnarray} 
\Phi=4.08 \times 10^{-2}\, \left( 
\frac{p_\mathrm{sat}}{1\,\mbox{torr}} \right) 
\,\, \left( \mu_\mathrm{} \frac{1\,\mbox{K}}{T} \right)^{1/2} 
~\mbox{g~cm$^{-2}$\,s$^{-1}$},
\label{eq:Phi}
\end{eqnarray}
where $p_\mathrm{sat}$\   is the phase-equilibrium vapor pressure 
at the surface temperature $T$, and  $\mu$ is the atomic weight of the 
water molecule.
There are different available prescriptions, usually valid
over different temperature ranges, for calculating 
the saturated vapor pressure over water ice surfaces
\cite[e.g., ][]{ Marti93,Buck96, Smithonian84}.
For the $T\ge 170$\,K range, we adopt the formula from 
\citet{Fanale84} 
\begin{equation}
p_\mathrm{sat} = 2.67 \times 10^{10} \exp
\left( - \frac{6141.667\,{\mbox{K}}}{T} \right) ~{\mbox{torr}},
\label{eq:psat}
\end{equation}
For $T < 170$\,K, we follow the experimentally derived results of
\citet{Mauersberger03} 
 \begin{equation}
p_\mathrm{sat} = 5.69 \times 10^{12} \exp
\left( - \frac{7043.51\,{\mbox{K}}}{T} \right) ~{\mbox{torr}},
\label{eq:psat_160}
\end{equation}
which converges with Eq.~\ref{eq:psat} at 170\,K.
The erosion rate of a spherical particle 
due to sublimation is then given by
 \begin{equation}
\dot{s}_\mathrm{sub}=-\frac{\dot{m}_\mathrm{sub}}{4 \pi \rho s^2}
=\frac{\eta\, \Phi}{\rho},
\label{eq:sdot_sub}
\end{equation}
where $s$ is the grain radius, $\dot{m}_\mathrm{sub}$ is the rate at which a grain loses mass,
$\rho$ is the  density  of the grain material, $\eta\,$ 
is the covering factor (fraction of the surface covered by 
sublimating material).
The sublimation lifetime of a grain then reads 
 \begin{equation}
t_\mathrm{sub}=\frac{s_0}{\dot{s}_\mathrm{sub}}=\frac{s_0\rho \eta\,}{\Phi},
\label{eq:t_sub}
\end{equation}
where $s_0$ is the initial radius  of the grain. 


Sublimation also affects the energy balance of a grain by
removing heat at the rate 
 \begin{equation}
E_\mathrm{sub}=4 \pi s^2 \eta\, \,L_\mathrm{s}  \Phi(T),
\label{eq:Esub}
\end{equation}
where $L_\mathrm{s}$ is the latent heat of sublimation.
Strictly speaking, $L_\mathrm{s}$ is a function of the grain temperature,
however the dependence is weak  
($2.6 \times 10^{10}$\,erg~g$^{-1}$ at 0\,K, $2.85\times 10^{10}$\,erg~g$^{-1}$ at
90\,--\,300\,K, see \citealt{Lien90})
and is not important for our calculations.  
We therefore adopt a value of $2.85 \times 10^{10}$\,erg~g$^{-1}$ 
through out
the computation.
At $T \ga 150$\,K, 
 $E_\mathrm{sub}$ is negligible compared with 
the energy loss rate due to thermal radiation (see Sect.~\ref{sec:results}).

Sublimation acts in the inner (warmer) part of a disc and  destroys icy
grains, or at least de-ices them, very efficiently. 
The radial dependence of 
the speed of evaporation process is very rapid, owing to $\Phi$ 
varying 
exponentially with $1/T$ which, in turn, depends on the distance 
from the star $R$. This allows us
to define the snowline location based on virtually any answer to the question
of what constitutes a ``fast enough'' evaporation rate: widely different definitions of
that rate all result in nearly the same locations of the snow line
(see Sect.~\ref{sec:subo}).

\subsection{Photosputtering}
\label{sec:sp}
Even outside the {\it thermal} evaporation zone, grains could be subject to 
destruction through  a mechanism that does not have nearly as sharp a spatial
boundary. Individual UV photons absorbed by a grain do not 
only dissociate water molecules but can cause molecule desorption from the
surface of a grain. This process is known as photosputtering or photodesorption.
Around the UV-bright stars the effect of 
photosputtering can be rather strong \citep{Art96} 
and the position of the actual snowline is no longer
determined  by sublimation alone. This is particularly true for
debris discs, which are optically thin, so that the outer
parts are not shielded from the UV radiation.

\begin{figure}[t]
\includegraphics[width=\columnwidth,clip]{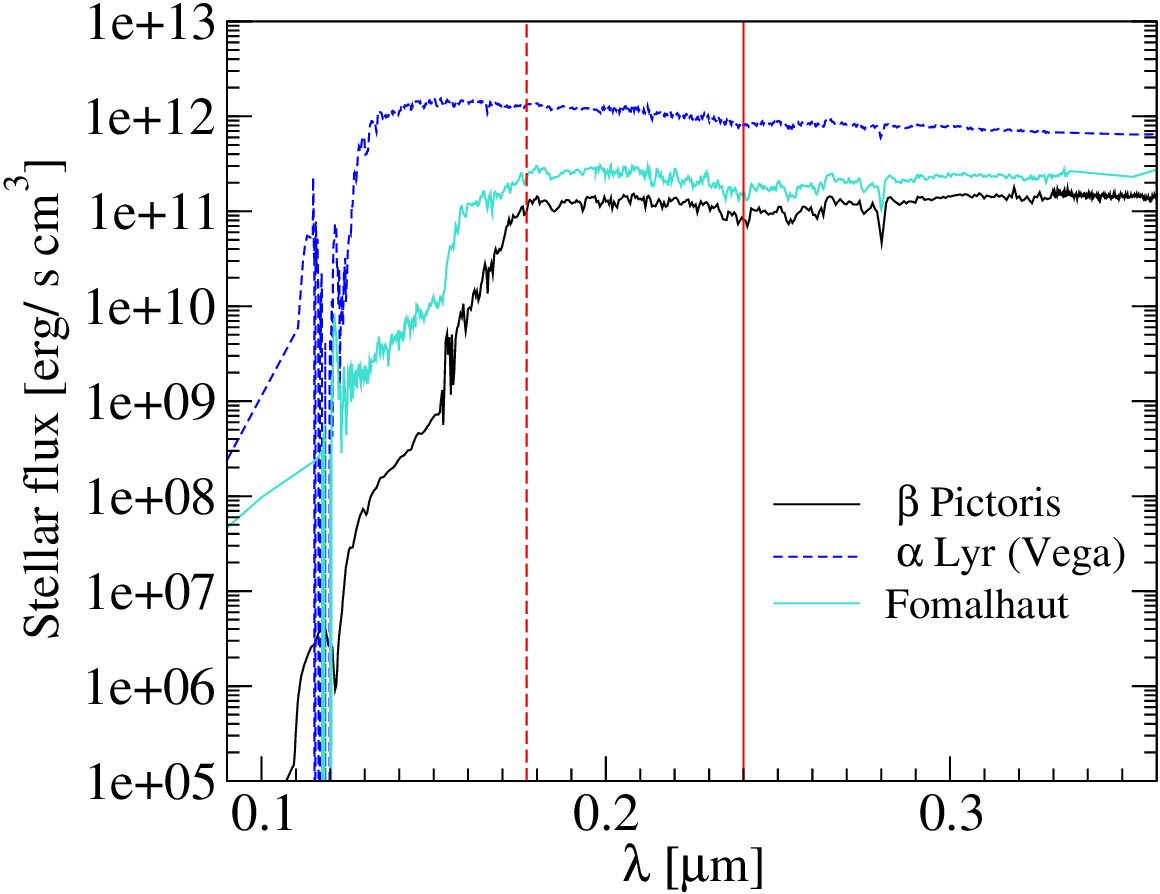}
\caption{Stellar fluxes, normalized to 1\,AU distance from the star.
The vertical solid line corresponds to 5.1\,eV (the energy of the O-H bond), 
the vertical dashed line marks 7\,eV (the energy at
 which the water ice absorption coefficients drop,
 Fig.~\ref{fig:H2O_Qabs}).}
\label{fig:stellar_flux}
\end{figure} 

The mechanism of photodesorption at the atomic level
has recently been investigated by \citet{Andersson06} with
their theoretical modeling of the interaction between  
8\,--\,9.5\,eV photons and icy surfaces.
First, a water molecule is dissociated by a UV photon.
After that, in some cases the OH$^{-}$ fragment can be directly desorbed 
or the molecule recombines and may leave the surface as a whole or it
can pass its momentum to one of the neighboring molecules, which can leave the 
surface if the  momentum is in the right direction.  
These authors estimated  the
desorption probability, called (photosputtering) yield $Y$,
and found it to lie between $4 \times 10^{-4}$ for amorphous ice
and $2 \times 10^{-3}$ for crystalline ice.
Yield estimates were also obtained by \citet{Westley95},
but this time from laboratory experiments, studying
UV sputtering of water ice by  Ly$_\alpha$ photons\,(10.2\,eV).
This study found that 
the yield is a weak function of the grain surface temperature
and of the irradiation dose.
At any given fluence (flux $\times$ exposure time),
$Y$ is slightly higher for warmer grains.
For a constant grain temperature, 
the yield increases rapidly with  irradiation dose and
reaches a plateau value for
doses higher than $\sim 3 \times 10^{18}$\,photons/cm$^2$ 
\citep[][]{Westley95}. 
This plateau value is 
on the order of few\,$\times\,10^{-3}$ at
$T$=35\,--\,100\,K. 
The obtained temperature dependence is relatively weak, the
extreme values being $Y=3\times 10^{-3}$ for $T\leq35$\,K and
$Y=7\times 10^{-3}$ for $T\geq100$\,K.

Direct comparison between the 
experimental results by \citet{Westley95} and those derived from
theoretical calculations by \citet{Andersson06} is not
straightforward.
On the one hand, the latter estimates were derived for
a fixed and very low temperature of 10\,K. On the other hand,
\citet{Westley95} data were obtained through laser beam
irradiation, which might have induced local point heating
and thus adding some sublimation to the pure
sputtering effect (N. J. Mason, private communication).
It is thus difficult to precisely derive
from these studies the exact dependence of
the yield on the energy of an absorbed photon
and the grain temperature.
However, it is worth noticing that both these independently
derived ranges
for $Y$ are comfortably of the same order of magnitude. 
We will thus make the simplified assumption that
$Y$ does not depend on temperature (which is reasonable, given
a weak temperature dependence found by \citealt{Westley95})
and is furthermore independent of the absorbed photon
wavelength
The constant $Y$ value we assume is $10^{-3}$, intermediate between
the lowest estimate of \citet{Andersson06} and the highest
estimate of \citet{Westley95}.

A crucial issue for the present problem is 
which energy range should be taken into account for UV
sputtering calculations. There is indeed no reason
why it should be restricted to the 8--11eV range explored
in the 2 aforementioned studies.
\citet{Dominik05}, in their study of an origin of gas-phase water 
in the surface layer of protoplanetary discs,   considered 
 photons
with energies down to 6\,eV and took a constant sputtering 
yield per incident photon. However,
since desorption is triggered by the initial dissociation of 
a water molecule (even if the molecule can recombine afterwards),
we adopt here a slightly lower threshold value of
5.1\,eV, which correspond to  the dissociation 
energy of the  O-H bond  \citep{Ruscic02}.\footnote{This energy still 
exceeds the surface binding energy of a lattice particle for water ice,
which  is on the order of 0.1\,--\,0.37\,eV \citep{Dijkstra03}.}
{{
An important point is that   only absorbed photos can play role in 
the photosputtering process. 
Photons considered by  \citet{Andersson06} and \citet{Westley95} 
belonged to the 8--11\,eV energy range. For such photons 
the absorption coefficient by 
water ice  is $\simeq 1$ (Fig.~\ref{fig:H2O_Qabs}), thus
the number of absorbed photons was very close to the number of incident photons.
For this reason, in this energy range it does not matter whether one considers
the number of insident or absorbed photons. However,   
water ice does not  efficiently absorb
UV photons in the $E_\mathrm{ph}\la 8$\,eV range, 
 and the difference between incident and absorbed fluxes is 
 significant. 
 Although for
  photons in the 5.1\,--\,7\,eV energy range $Q^{\prime}_\mathrm{abs} \ll 1$, 
  the effect of such ``low energy'' photons cannot be neglected,
because many stars radiate significantly more in the
5\,--\,7\,eV range than above 8\,eV (see the  $\beta$~Pic
case in Fig.~\ref{fig:stellar_flux}). However, the excact value of the lower
 $E_\mathrm{ph}$ boundary is not critical, provided it is below $\sim 6.5$\,eV.

For the upper boundary of the energy of sputtering photons,
we follow \citet{Dominik05} and take 13.6\,eV.
The exact value for this upper limit is not important 
as the amount of radiation beyond 11\,eV
can be neglected for most stars (see Fig.~\ref{fig:stellar_flux})

}}

For an icy grain, the erosion rate due to UV sputtering is calculated then as
 \begin{equation}
\dot{s}_\mathrm{sp}=-\frac{\eta\, m_\mathrm{(H_2O)} Y N_\mathrm{abs}}{4\rho}, 
\label{eq:sdot_sp}
\end{equation}
where $\eta $ is the fraction of surface covered by ice, 
$ m_\mathrm{(H_2O)}=3 \times 10^{-23}$\,g is the mass of a 
water molecule 
and $N_\mathrm{abs}$ is the number of absorbed photons 
 \begin{equation}
N_\mathrm{abs}=\int_{\lambda_{\mathrm{min}}}^{\lambda_{\mathrm{max}}} 
\frac{F(R,\lambda)}{hc/\lambda}\, Q^{\prime}_\mathrm{abs}(\lambda)\, d\lambda, 
\label{eq:Nabs}
\end{equation}
in which $F(\lambda,R)$ is the incident stellar 
flux at the position $R$ of the grain,
$\lambda_{\mathrm{min}}=0.091\,\mu$m (13.6\,eV), 
$\lambda_{\mathrm{max}}=0.24\,\mu$m (5.1\,eV), $Q^{\prime}_\mathrm{abs}(\lambda)$
is the absorption coefficient for pure water ice surfaces 
(Fig.~\ref{fig:H2O_Qabs}).

\subsection{Gas production \label{sec:gas}}

Sublimation and photosputtering of ice produce gas.
A significant fraction of gas in debris discs
 could be produced by the solids rather than 
 be leftovers of the primordial dust component.
 This opens a possibility to 
 deduce dust   composition 
 from abundances of various species in the gas component
 (if we know the parameters of the desorption processes).
 
 In the case of icy grains, gas production rate due to
 UV sputtering can be calculated as 
%
\begin{eqnarray}
\label{Mdot00}
\dot{M}_{\mathrm{H_2O}} & = & \int\!\! \int\!\! \int \eta\, 4\pi \rho s^2 
\dot{s}_{0} \left( \frac{1\,\mbox{AU}}{R} \right)^2 
dn(s)\, dz\, 2\pi R dR \nonumber \\
 &= & 
8 \pi \rho \dot{s}_{0} \eta\,(1\,\mbox{AU})^2 
\int \left( \frac{\tau_{\perp}}{R} \right)dR,
 \end{eqnarray}
%
where $\dot{s}_{0}$ is the sputtering rate at 1\,AU from the star,
$dn(s)$ is the number density of grains of size $s$, $z$ is the height 
above the midplane, and $\tau_{\perp}(R) $ is the normal geometrical 
optical thickness of the disc.
 If a disc  has a constant opening angle, i.e. a constant $H/R$ ratio, 
 where $H$ is the height  scale of the disc, then
 the above equation can be rewritten as  
\begin{equation}
\label{eq:Mdot}
\dot{M}_{\mathrm{H_2O}} = 8 \pi \rho \dot{s}_{0} \eta \,(1\,\mbox{AU})^2 
\tau_{\|} \left( \frac{H}{R} \right),
 \end{equation}
 where $\tau_{\|}$ is the geometrical optical thickness of 
 solids in the disc along a radius in the midplane.

\subsection{Grain temperature} 

As we have discussed, both sublimation and photosputtering 
erosion rates depend on grain temperature.
This dependence is weak for UV sputtering but temperature
obviously plays a crucial role for sublimation
and should thus be accurately computed.
The grain temperature is calculated from the energy balance
equation, in which the radiative heating by the central star
is balanced by energy loss to thermal re-radiation (assuming uniform temperature
of the surface) and sublimation
 \begin{equation}
\begin{array}{l}
\left(\frac{R_*}{R}\right)^2
\int_0^{\infty} Q_\mathrm{abs}(s,\lambda) F_*(\lambda)\, d\lambda= \\
\\
4\left[\int_0^{\infty} Q_\mathrm{abs}(s,\lambda)\,\pi B_{\lambda}(T)\, d\lambda +
 \eta\, \,L_\mathrm{s}  \Phi(T)\right],
\label{eq:T}
\end{array}
\end{equation}
where $r$ is the distance from the star,
$R_*$ is the stellar radius,
 $F_*$ is the stellar flux at $R_*$, 
$B_{\lambda}(T)$ is the Planck function at equilibrium temperature $T$,
$s$ is the grain radius, and   $Q_\mathrm{abs}$
is the absorption coefficient for the whole grain.
Notice that in Eq.~\ref{eq:T}
we use the ``true''  $Q_\mathrm{abs}$ obtained for a given grain composition
whereas
in Eq.~\ref{eq:Nabs} we use the absorption coefficient $Q^\prime$
of pure water ice. 
The reason for this is that the grain temperature depends on 
absorption by the whole grain, while 
the UV sputtering on absorption by water molecules near the the surface.
 
The absorption coefficient  $Q_\mathrm{abs}$ is a function of   
grain size and chemical composition \citep[e.g.][]{BurnsLamy79}.
It is here calculated using
the Mie theory code developed by \citet{Art88}. 
Effective medium theory is used to obtain effective dielectric constants
or equivalently refraction indices, based on material component's 
properties and volume mixing ratios. $Q_\mathrm{abs}$
plays a crucial role in calculations of the  grain temperature   
and, as a consequence, in determination of the $T(R)$ 
dependence. A previously popular piecewise power-law ansatz 
($Q_\mathrm{abs}=1$ for $\lambda \le s$,
 $Q_\mathrm{abs}=s/\lambda$ otherwise)
is not suitable for calculating the  temperature of the predominantly 
icy grains, because it gives a significant overestimate of the temperature and, 
as a consequence, much overestimated radius of the ice sublimation boundary.

\subsection{Icy grain erosion and dynamics} \label{sec:dynamics}

In our model we not only estimate sputtering and sublimation rates,
we also compute the effect these mechanisms have on  
grain dynamics in a collisionless, gas-free disc. We follow the evolution
of an icy grain taking into account the above two erosion modes,
and the resultant change of the size-dependent 
radiation pressure coefficient $\beta$ (ratio of radiation
pressure to gravity). 
The equation of motion of a grain reads

\begin{eqnarray} 
\label{eq:F}
m\,\frac{d^2\vec{R}}{dt^2}= 
 -\, \frac{GMm}{R^3}\,\vec{R}(1-\beta) +\vec{F}_\mathrm{PR},
\end{eqnarray} 
where $m$ and $\vec{R}$ are the mass and position of the test particle,
$G$ is the gravitational constant, and $M$ is
the stellar mass.
$\vec{F}_\mathrm{PR}$ is the  Poynting-Robertson (PR)
drag:
\begin{eqnarray} 
\label{eq:PR}
\vec{F}_\mathrm{PR} = - \beta\, \frac{GMm}{R^2} 
\left(\frac{v_r}{c}\frac{\vec{R}}{R}+\frac{\vec{v}}{c} \right),
\end{eqnarray} 
where $\vec{v}$ is the velocity vector of the grain,
$v_r$ is its radial component, and $c$ is the speed of light.

\begin{figure}[t]
\includegraphics[width=\columnwidth,clip]{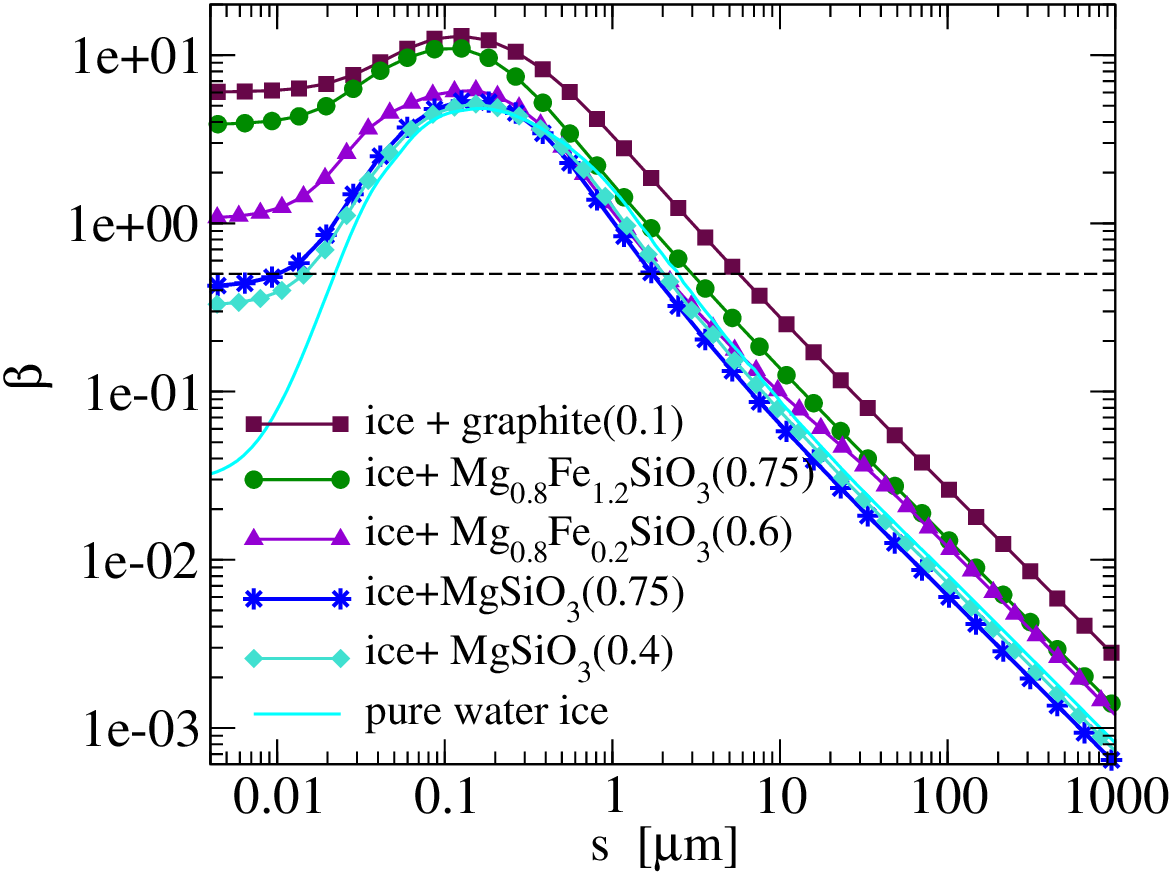}
\caption{Ratio of the radiation pressure force to the gravitational
force vs. grain size for different grain compositions
around an A5V star (like $\beta$~Pic).
Ice means amorphous water ice. The numbers in parentheses 
indicate the volume fraction of the corresponding elements.}
\label{fig:beta}
\end{figure} 

The coefficient $\beta$ depends on the
grain size, chemical composition and stellar luminosity (more precisely, 
luminosity to mass ratio $L/M$).
In our code,  $\beta$  is calculated using the routines
from \citet{Art88}, and displayed in Fig.~\ref{fig:beta}. For sizes above
a certain threshold value ($s\ga0.3\,\mu$m for the case displayed here),
$\beta(s)$ is well approximated by the geometrical optics approximation 
 \begin{equation}
 \beta=0.5\,\frac{s_{1/2}}{s},
 \label{eq:beta}
\end{equation}
where $s_{1/2}$ is the grain radius for which $\beta=0.5$. 
In this approximation, the gradual decrease in grain radius $s$ due to 
erosion will cause slow changes in $\beta$ at a rate
 \begin{equation}
 \dot{\beta}=- \,0.5\,\frac{s_{1/2}}{s^2}\,\dot{s}, 
 \label{eq:beta_dot}
\end{equation}
where $\dot{s}=\dot{s}_\mathrm{sub}+\dot{s}_\mathrm{sp}$.
Since $\dot{s} <0$, $\dot{\beta}>0$, 
at least for grains bigger than few tenths of a micrometer
(which is, incidentally, the wavelength of the peak power of radiation 
exerting the pressure). 
As can be seen from Eq.~\ref{eq:F}, such gradual changes in $ \beta$ 
will increase the outward radiation pressure, which will tend
to push the particle outwards, but at the same time, 
also increase the PR drag which tends to move particles inwards.
For a grain on a Keplerian orbit of semi-major axis $a$,
one can show that, for small eccentricities, the rate at which
$a$ changes reads
\begin{eqnarray} 
\label{eq:adot}
\frac{\dot{a}}{a}=\frac{\dot{\beta}}{1-\beta}-
\frac{2\,\beta\, G M}{(1-\beta)\, a^2c}=
\frac{1}{1-\beta} \left( \dot{\beta} - t_\mathrm{PR}^{-1}\right),
\end{eqnarray} 
where $ t_\mathrm{PR}=a^2 c/(2 \beta\, GM)$ is the Poynting-Robertson 
time. 
If $\dot{\beta}> t_\mathrm{PR}^{-1}$, i.e., if the  erosion rate is
larger than about $ 2\,{s^2}/({s_{1/2} \,t_\mathrm{PR}})$,
or the erosion timescale $s/dot{s}$ is shorter than 
 $ (s_{1/2}/2s) \,t_\mathrm{PR}$,
 then  {\it{PR inward}}
 migration is replaced by the {\it{outward}} erosional migration.

\section{Results}  \label{sec:results}

We now present typical grain behaviors 
for different representative stellar types. For
sake of clarity, we first consider a nominal case, for which 
we present the results  in detail, 
to fully describe the mechanisms at work.
For this reference case, we consider the archetypal
$\beta$~Pic system. Results for other stellar
types are presented in the following subsections.

\subsection{A detailed example: \bp}  \label{sec:bp}

$\beta$~Pic is an A5V main-sequence star  
surrounded by an optically thin
debris disc, which was directly imaged  for the first time
more than 20 years ago \citep{Smith84}.
We take this system as our reference case, since it is still by
far the most extensively studied and best known debris disc system.
Countless observational  
and theoretical studies have been dedicated to this system 
 \citep[e.g., the most recent ones: ][]{ Chen07, Golimowski06, Tamura06,
Galland06, Roberge06, Fernandez06, Telesco05,
Thebault05, Brandeker04, Karmann03, Au01}.
The disc has a wide radial extension (up to 1000\,AU), but
most of the detected dust is thought to reside in a
relatively narrow region between 80 and 120\,AU \citep[see for example
the fit of the scattered and thermal light profiles derived by][]{Au01}.
In a few studies 
water ice was considered as a possible grain component.
As was already mentioned in Sect.~\ref{sec:introduction},
\citet[][]{Golimowski06} considered the presence of icy mantles as a
possible explanation for the observed color changes.
Likewise, \citet{Tamura06} needed an icy component 
to reproduce their high resolution near-infrared and previous 
polarimetry observations in terms of dust scattering.
In the same spirit,
\citet{Au01} assumed that ice might be present beyond 
the sublimation boundary. 
However, as for most debris disc systems, there is no
``direct'' unambiguous observational proof for the  presence
of icy grains. 
So far, the existence of ice is thus still conjectural. 

We address this issue by careful modeling of the criteria 
for the survival of water
ice. The sublimation boundary alone is first considered 
in a ``static'' case
when grains are assumed to stay at a fixed distance from the star. 
The photosputtering rate and corresponding survival times 
in the same ``static'' 
assumption are estimated afterwards. Finally, 
the coupled effect of these mechanisms is investigated when
the grain dynamical evolution is taken into account . 
In doing so,  we imagine ice
to be predominantly in H$_2$O grain mantle, thus  covering the surface
($\eta=1$). This maximizes the erosional and dynamical evolution.

\subsubsection{Sublimation boundary}
\label{sec:subo}

 \begin{figure}[t]
\includegraphics[width=\columnwidth,clip]{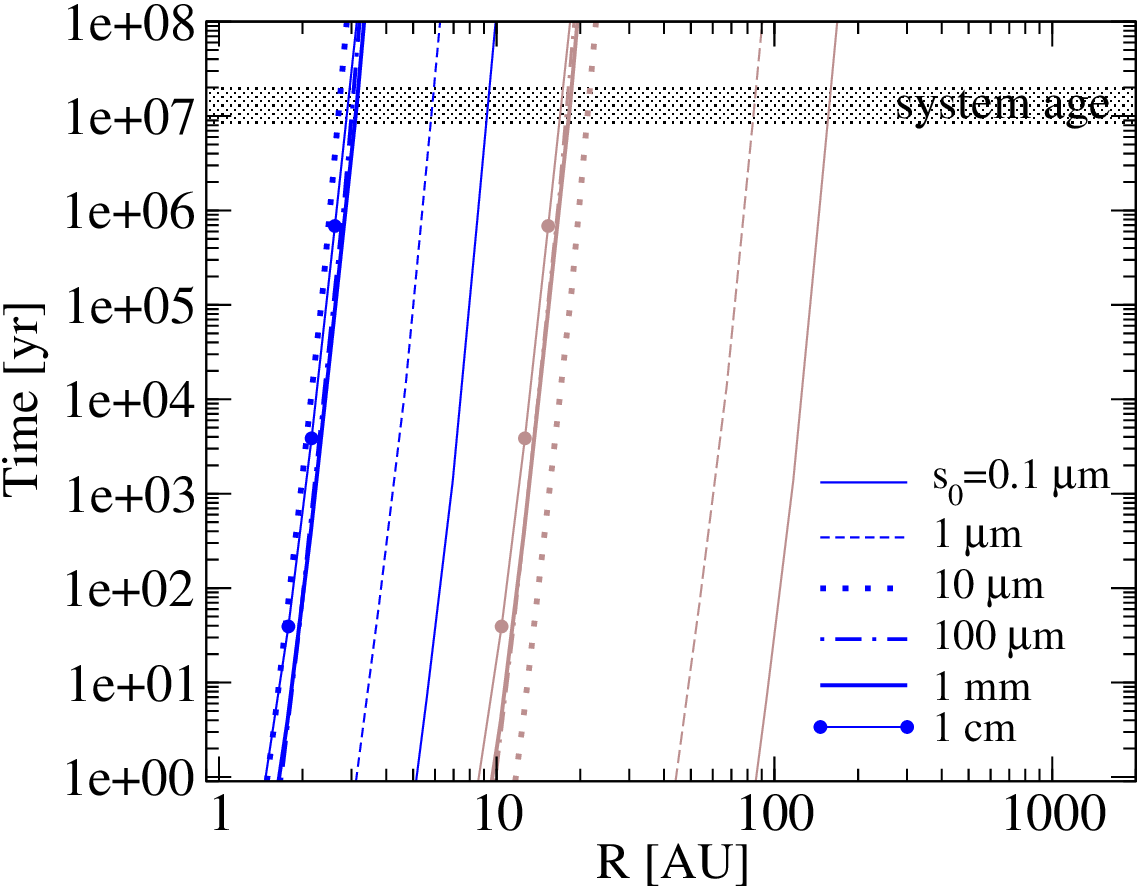}
\caption{Sublimation timescales for grains of different sizes 
and compositions, in the ``static'' assumption where
grains are assumed to stay at a fixed radial distance.
\emph{Dark blue lines:} pure water ice grains, which are 
the coldest ones.  \emph{Light brown lines:}
icy grains with 10\,\% of graphite inclusions, which 
represent some of the ``hottest'' grains.  
These two choices of grain compositions are  
not the most representative from the point of view 
of grain chemistry. However, they  bracket 
temperature values for a grain of a certain size
at a give distance from the star.  }
\label{fig:tsub}
\end{figure}

\begin{figure}[t]
\includegraphics[width=\columnwidth,clip]{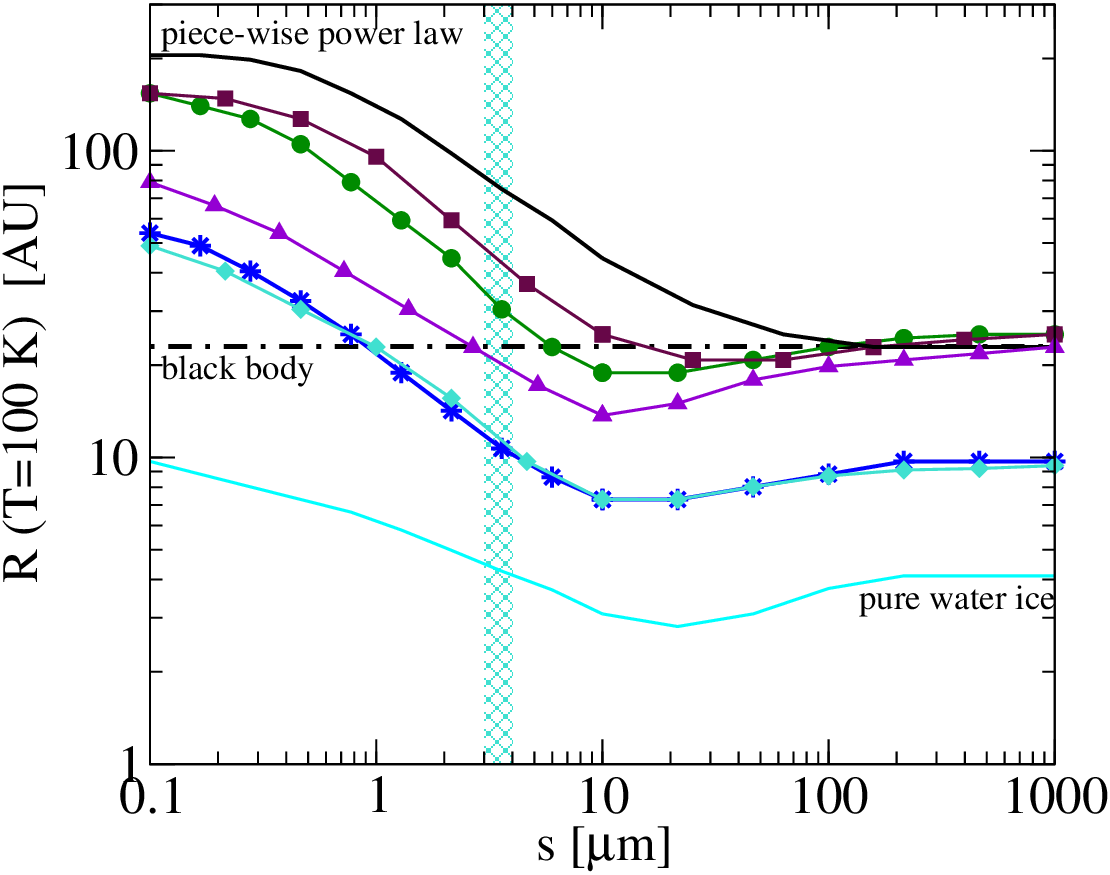}
\caption{Isotherms ($T=100$\,K) as a function of grain size for different 
grain compositions. These lines provide the outer limit for ice sublimation
boundary. This plot clearly shows  how the sublimation boundary depends
 on the grain sizes
and chemical compositions. The horizontal line is for 
the blackbody approximation. The solid black line is for the artificial piece-wise power low dependence
(  $Q_\mathrm{abs}=1$ for $\lambda \le s$,
 $Q_\mathrm{abs}=s/\lambda$ otherwise$) $.
 All other lines are the same as in 
Fig.~\ref{fig:beta}
}
\label{fig:R100K}
\end{figure} 

\begin{figure}[t]
\includegraphics[width=\columnwidth,clip]{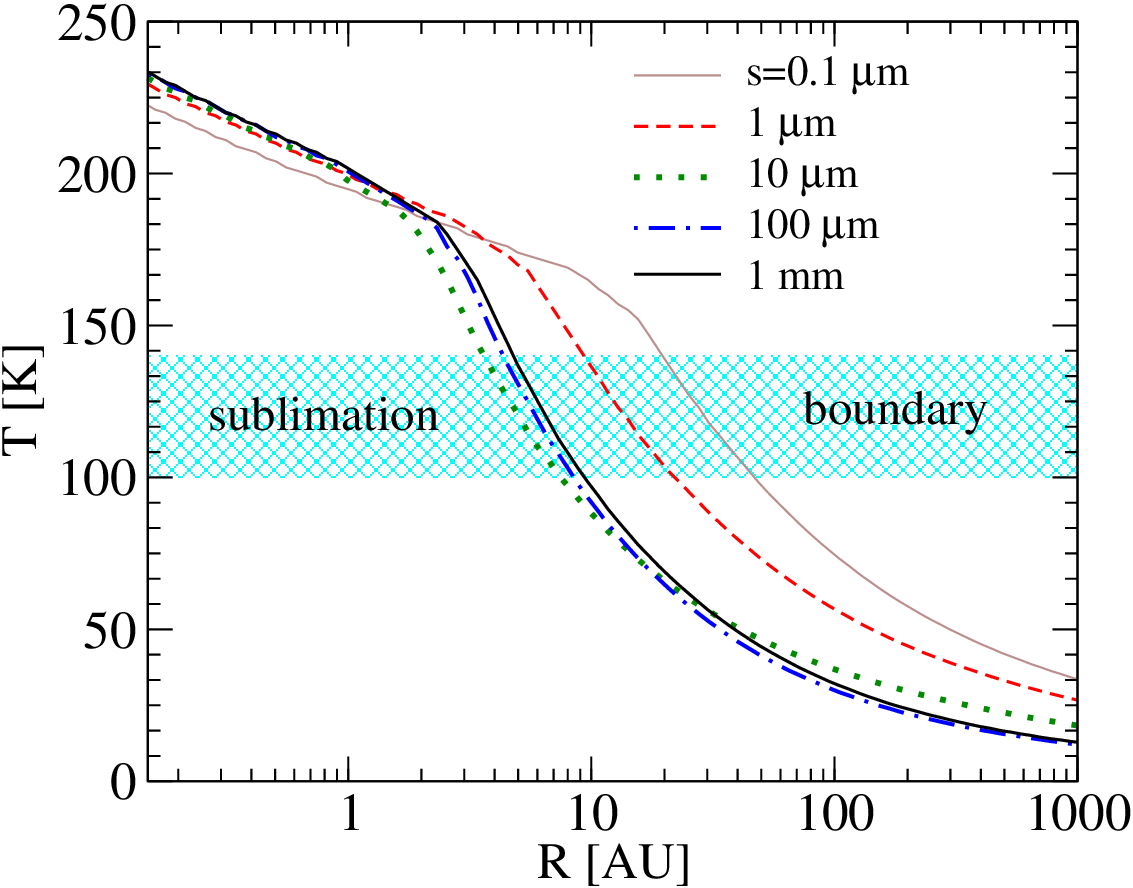}
\caption{Temperatures of dirty ice grains (60\% of water ice + 
40\% of MgSiO$_3$) as a function 
of the distance from the star (for the \bp~case). The sublimation boundary 
lies somewhere in the crosshatched region. The grain radii are 
indicated on the plot.
 }
\label{fig:T_2_20}
\end{figure}

The sublimation boundary can be defined as 
the largest distance from the star at which the sublimation time 
of predominantly icy particles is shorter 
than any other relevant time scale \citep{Art97}.
We take here a very conservative range and consider
the 2 extreme timescales given by the orbital period $t_\mathrm{orb}$, in
the 1\,--\,100\,yr range, and the system's age $t_\mathrm{age}$, 
i.e. 8\,--\,20\,Myr \citep{zuc01,dif04}.
The crucial parameter is  the temperature, since
$t_\mathrm{sub}\propto \exp(-1/T)$ (Eqs.~\ref{eq:Phi} to \ref{eq:t_sub}).
The temperature dependence is in fact so sharp that
the temperature for which $t_{\mathrm{sub}}(T)=t_\mathrm{orb}$, i.e.,
$\simeq 140$\,K,
is relatively close to the one for which $t_{\mathrm{sub}}(T)=t_\mathrm{age}$,
i.e. $\simeq 100$\,K. As a consequence, we
shall as a first approximation neglect the timescale issue and
conservatively consider that the sublimation boundary
lies somewhere in this 100\,--\,140\,K range.
The temperature of a given grain depends schematically on 3 parameters:
{{its size, chemical composition}} and distance to the star.
This is illustrated in Fig.~\ref{fig:R100K} where
100\,K-isotherms, giving the outer limit of the sublimation boundary,
are plotted for different grain compositions as a function of grain size.  
If we consider a reasonable intermediate compositional case of
dirty ice grains, we see that the sublimation boundary lies
somewhere between 5--10\,AU for large mm-particles and
20--50\,AU for hotter submicron-scale grains (Fig.~\ref{fig:T_2_20}).
{ {For compact grains  these values 
are all far inside
the location of the color change observed by 
\citet{Golimowski06} at $\simeq120$\,AU. Only very small 0.1\,$\mu m$
grains of special chemical compositions have  an outer
sublimation boundary approaching 120\,AU (Fig.~\ref{fig:R100K}), but
those grains should be blown out by radiation ($\beta > 1$).
Thus, in the case of compact grains, our results rule out
sublimation as a simple explanation for a hypothetical 
compositional transition at $\sim$120\,AU from the star.
However, if grains are highly porous, as e.g. was introduced in \citet{Li98b},
and consist of many individual submicron grains, the result can be different,
since  such aggregates optically behave like  invidual 
  submicron grains.}} 

{{
\subsubsection{Recondensation}
Here we  justify that the recondensation rate is
very small and can be safely neglected in  a debris disc of 
a late type, nearly gas free system
 like, e.g., $\beta$~Pic.
 
 The condensation rate can be computed with the help of
 Eq.~\ref{eq:Phi}  if instead of $p_{sat}$, 
the actual
water vapor pressure above the ice surface $p$ is substituted. 
In the ideal gas approximation, $p=kn_\mathrm{H_2O}T_\mathrm{g}$ 
where $k$ is the Boltzmann constant,
$n_\mathrm{H_2O}$ is the water molecule volume density in gas phase, $T_\mathrm{g}$ is the 
gas temperature. If the gas and the grain surface are in thermal
equilibrium $T_\mathrm{g}=T$, the condensation rate, $\Phi_{\mathrm{c}}$,
 can be calculated as
\begin{eqnarray} 
\Phi_{\mathrm{c}}=1.8 \times 10^{-26}\, \left( 
\frac{n_\mathrm{H_2O}}{1\,\mbox{cm$^{-3}$}} \right) 
\,\, \left(  \frac{T}{1\,\mbox{K}} \right)^{1/2} 
~\mbox{g~cm $^{-2}$\,s$^{-1}$}.
\label{eq:Phi_c}
\end{eqnarray}
To estimate $\Phi_{\mathrm{c}}$ we need to know $n_\mathrm{H_2O}$. 
The  upper limit can be obtained from the H
distribution given in \citet{Brandeker04} for the solar composition
case under an oversimplified assumption that all oxygen atoms are
locked in water vapor. The assumption is not realistic, since some
fraction of oxygen is present in atomic phase, some is locked in CO
and O$_2$, but it gives an easy-to-estimate upper limit. This
estimate\footnote{Detailed calculations of the
ionization disk structure and the equilibrium chemistry network, show that  
for the solar
composition case  from  \citet{Brandeker04} 
 $n_\mathrm{H_2O}\la 10^{-6}$\,cm$^{-3}$ 
and in
the metal depleted case  $< 0.1$\,cm$^{-3}$ 
(Liseau, R. private communication)}  gives us
$n_\mathrm{H_2O} < 1$\,cm$^{-3}$,
  resulting in  $\Phi_{\mathrm{c}} \la 10^{25}$g~cm $^{-2}$\,s$^{-1}$ 
 and a recondensation rate that exceeds the sublimation
 rate at $T\la 85$\,K. Thus recondensation dominates sublimation
  in the outer regions of the disk, but the time scale, even for the case of 
  orders-of-magnitude overestimated  H$_2$O vapor densities, is so high (it would take
  about $ 10^{12}$\,yr to accumulate 1\,$\mu$m  layer of ice on a grain surface)
   that this
  process can be safely ignored in the frame of debris disk systems.

}}

\subsubsection{UV sputtering\label{sec:uvsputt}} 

\begin{table} 
\begin{center}
\caption{Sputtering rate by UV photons at 1\,AU from \bp~  
for different grain radii.} 
\label{tab:sdot_sp}
\begin{tabular}{cc}
\hline\hline
$s$\,($\mu$m) & $\dot{s}_\mathrm{sp}$\,($\mu$m~yr$^{-1}$) \\ 
\hline
0.1 & 1.3  \\
0.3 & 1.8  \\
 1&	 2.35 \\
3 &	3.23 \\
7 &	4.14 \\
$\ge $ 10 &	4.45 \\
\hline
\end{tabular}
\end{center}
\end{table}

Since the disc is optically thin\footnote{Neither the amount of dust 
nor the amount of hydrogen 
(the  upper limits:  $N$(HI)$\la 5 \times 10^{19}$\,cm$^{-2}$ 
\citep{Freudling95}, $N$(H$_2$)$\la 3 \times 10^{18}$\,cm$^{-2}$ 
 \citep{Lecavelier01}) is enough to effectively shield the UV 
radiation}, UV sputtering occurs even in the outermost regions.
Taking into account the geometrical dilution of the 
stellar UV radiation and adopting our conservative and 
temperature-independent value of $Y=10^{-3}$ (see section
\ref{sec:sp}), we obtain for the UV sputtering rate 
\begin{equation}
\dot{s}_\mathrm{sp} \approx  - 4.45 \left(
\frac{1\,\mbox{AU}}{R} \right) ^2\,\mu\mbox{m yr}^{-1}.
\label{eq:sbp}
\end{equation}

\begin{figure}[t]
\includegraphics[width=\columnwidth,clip]{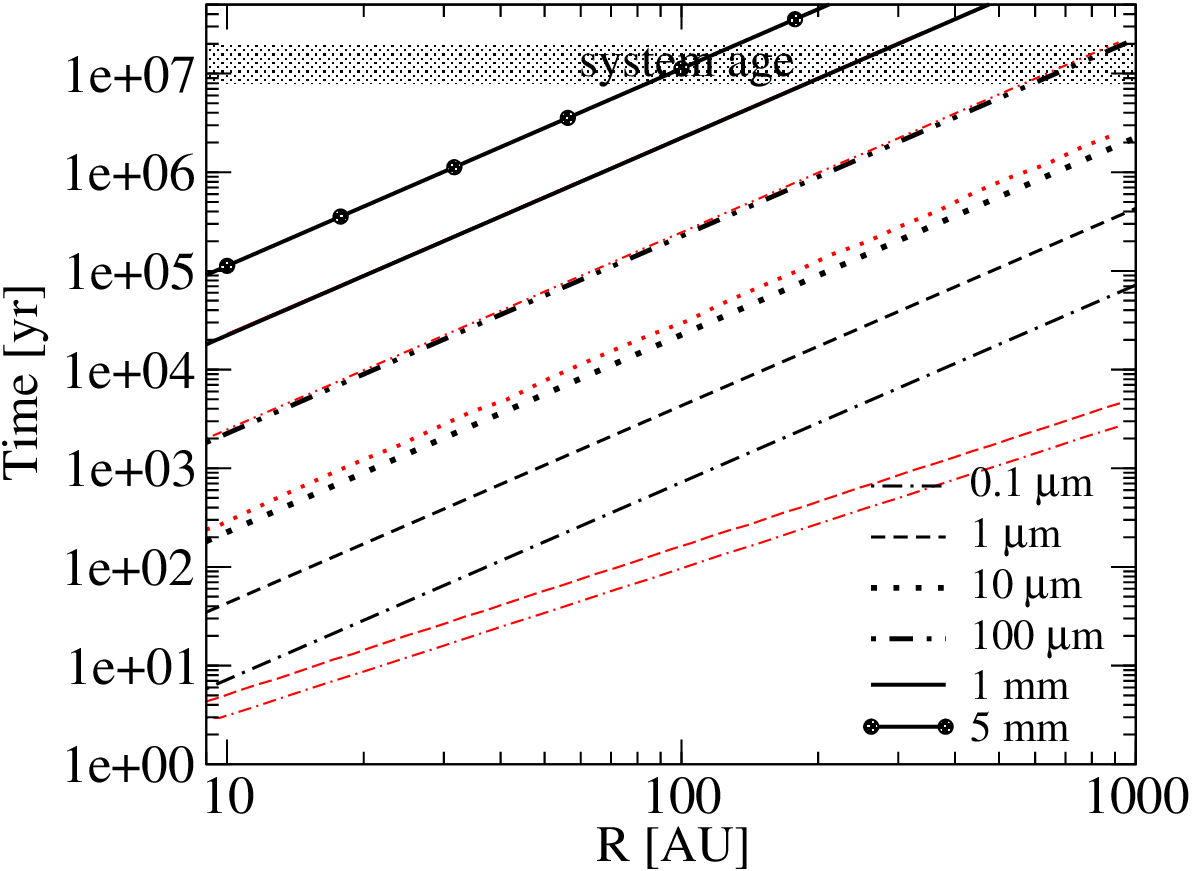}
\caption{
{ \it {Thick black lines:}}
UV sputtering destruction timescales for 
icy grains at different distances from  \bp~ central star
(for $Y(T)=10^{-3}$). 
The calculations are done for the ``static'' case, i.e.
grains are assumed to stay at the same distance from the 
star. { \it {Thin red lines:}} Removal timescales $t_\mathrm{rem}$
derived from Fig.~\ref{fig:tremov} when taking into account
grain dynamics.
 }
\label{fig:time_sp}
\end{figure}

The front coefficient is a weak function of 
grain size (Table~\ref{tab:sdot_sp}), 
since the number of absorbed photons depends
on $Q_\mathrm{abs}^{\prime}$.
The corresponding sputtering destruction
timescale $t_\mathrm{sp}=s_0/\dot{s}_\mathrm{sp}$ has been
plotted in Fig.~\ref{fig:time_sp}.
A first important result is that the dependence with
$R$, while being significant, is much less pronounced than for the sublimation
case (Fig.~\ref{fig:tsub}). There is no longer an abrupt transition
between the non--eroding and eroding regions, and it is difficult
to define a ``snowline'' due to sputtering.
A consequence of this result is that sputtering is 
unlikely to explain
sharp observed changes such as the color transition
at 120\,AU.

Another important result is that no ``dust'' grain (i.e. in
the $\la 1$--5\,mm range) can survive in the inner $R \la200$\,AU
region over the system's age. Even in the outer regions,
only relatively large grains,
in the sub to millimetre range, can survive.
We shall consider, as a first approximation, that the minimum
size for a grain which can retain its icy component over the system's age
in the $\simeq$80--120\,AU region where most of the $\beta$~Pic dust has
been observed is $\simeq 5$\,mm.

\subsubsection{Icy grain dynamics} \label{sec:dynamics_bp}

\begin{figure}
\includegraphics[width=0.7\columnwidth,angle=0]{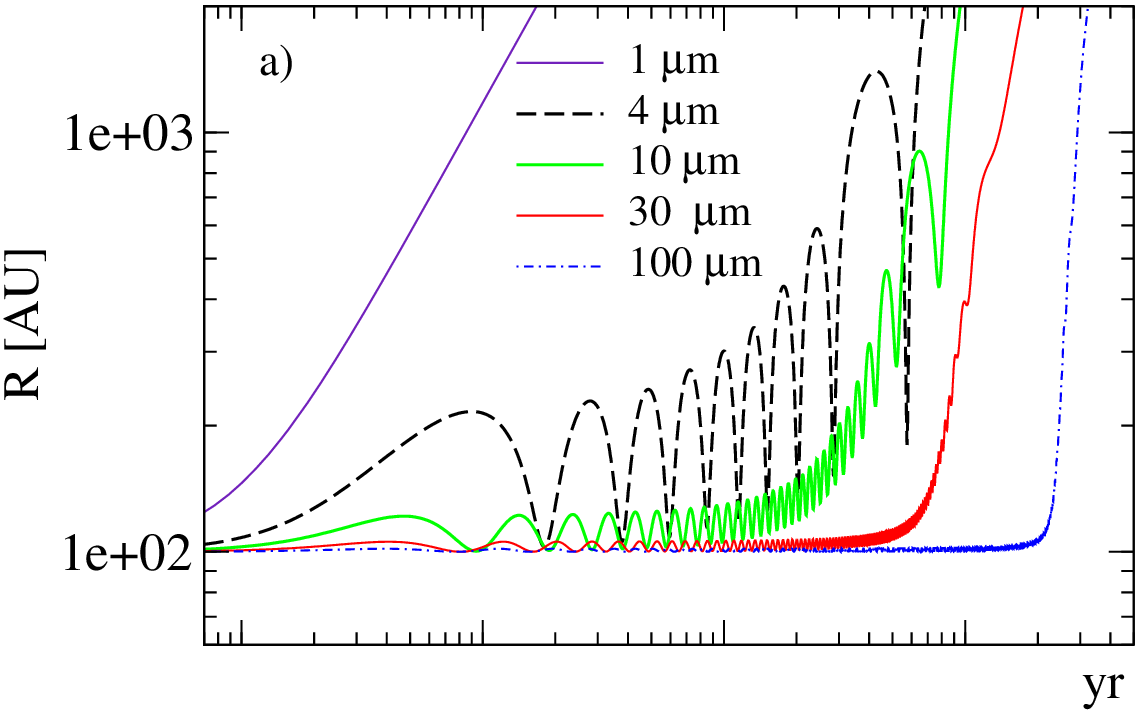}
\includegraphics[width=0.7\columnwidth,angle=0]{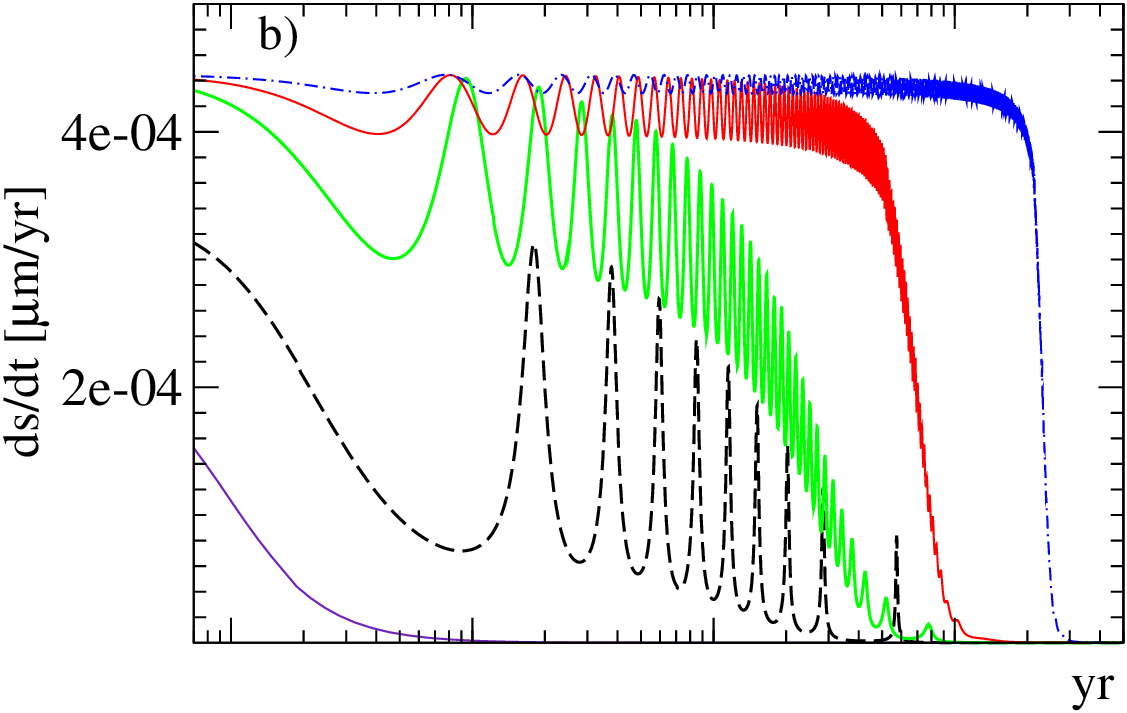}
\includegraphics[width=0.7\columnwidth,angle=0]{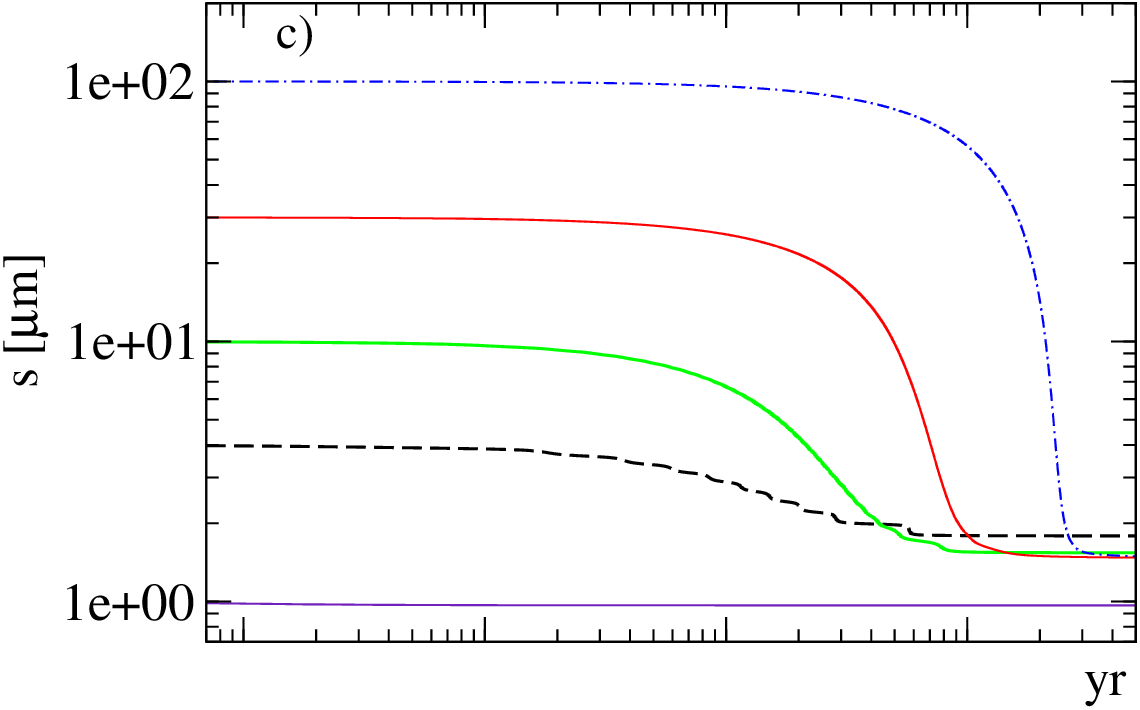}
\includegraphics[width=0.7\columnwidth,angle=0]{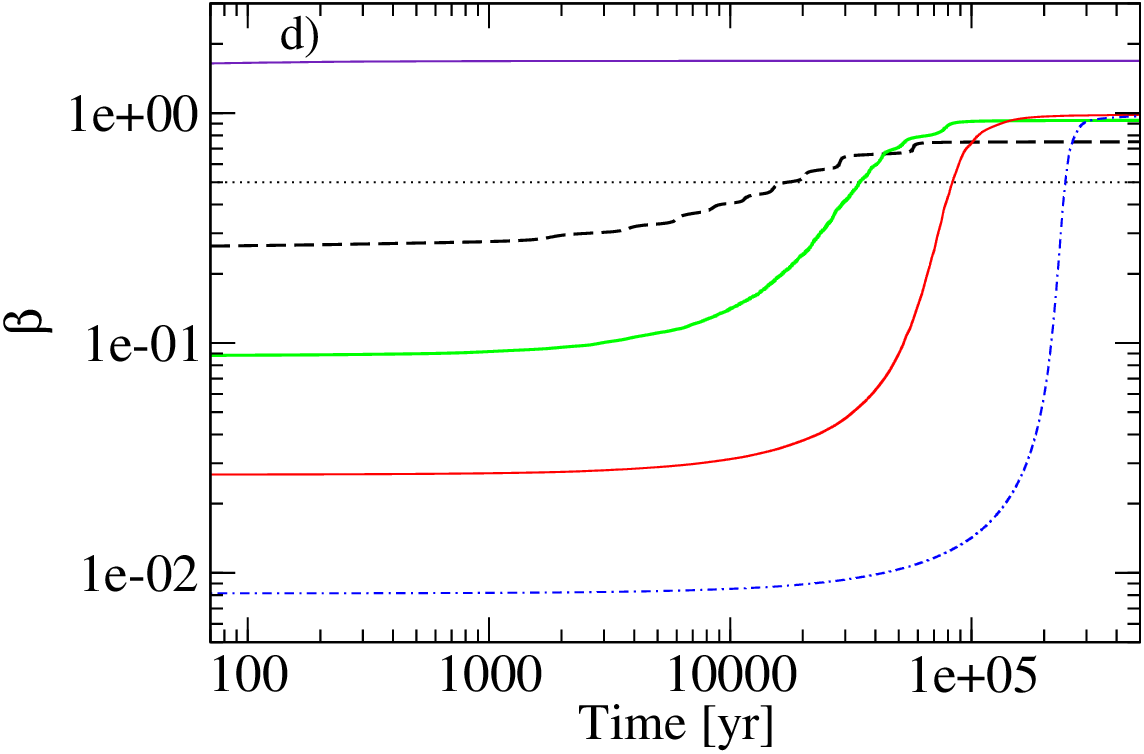}
\caption{Dynamics of UV sputtered icy grains, initially placed 
at 100\,AU from the star at a keplerian velocities. 
}
\label{fig:grain_dynamics}
\end{figure}


All the time scales previously discussed were given for the
``static'' case, when a grain was assumed to stay at a given distance from the 
star. This is of course a simplification of the behavior of a ``real''
particle. Indeed, as a grain gets eroded, be it by sublimation
or UV sputtering, its response to radiation pressure (measured
by its $\beta$ parameter) changes.
As has been shown in Sect.~\ref{sec:dynamics}, this has consequences
on the grain's orbital parameters, leading to outward or inward migration
depending if the ratio between the size erosion timescale is
shorter or longer than the typical Poynting-Robertson timescale 
$t_{\mathrm{PR}}$.

The full dynamical effect of size erosion has been taken
into account in a batch of simulations where test particles motions
are obtained by integrating Eq.~\ref{eq:F} (Fig.~\ref{fig:grain_dynamics}).
Our results show that for all initial particle sizes considered,
we are here always in the outward migration regime. All particles
released at 100\,AU reach 1000\,AU on a timescale comprised between
a few $10^{4}$\,yr for the smallest considered bound grains
 ($4\,\mu$m)
and a few $10^{5}$\,yr for large submillimetre grains. 
An interesting consequence of these strong outward migrations is
that no grain ever gets fully eroded to $s=0$. In fact, once 
grains enter the outbound orbit regime ($\beta\ga 0.5$), the outward
migration rate increases very sharply and gets much higher than the
size erosion rate. This imbalance between migration and
erosion rates is self-amplifying: as the grain
gets ejected its erosion rate rapidly decreases 
(since $t_{\mathrm{sp}} \propto R^{2}$).
Eventually, the grain gets ejected from the
system before it has time to be fully eroded, and its size
no longer evolves once the grain has passed beyond $\simeq 1000$\,AU
and reaches an asymptotic value (see Fig.~\ref{fig:grain_dynamics}c).
Interestingly, this asymptotic value, which is in the 1--2\,$\mu$m range,
is relatively independent of the grain's initial size $s_0$,
the only difference being that it is reached faster for initially
smaller particles.

\begin{figure}[t]
\includegraphics[width=\columnwidth,clip]{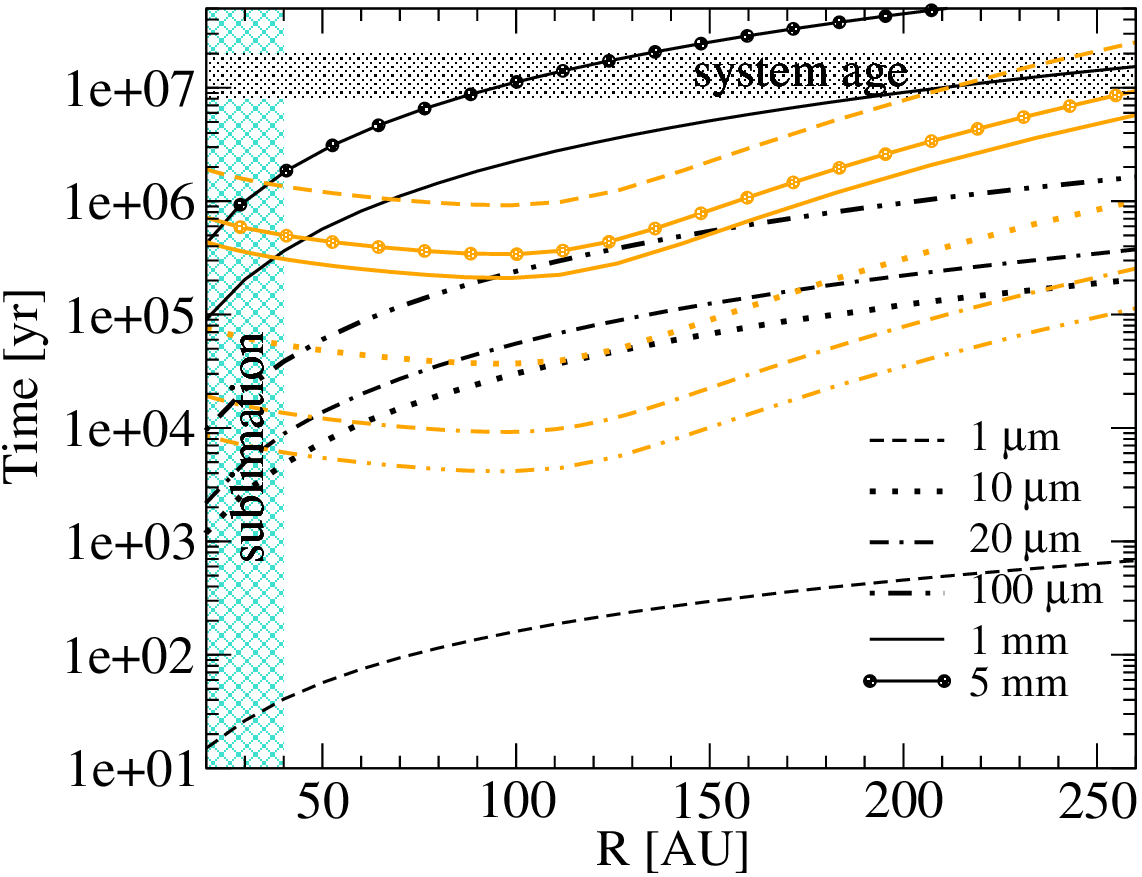}
\caption{Sputtering removal timescales $t_\mathrm{rem}$
and collisional lifetimes $t_\mathrm{coll}$
for icy grains of different sizes. 
All $t_\mathrm{rem}$ curves appear as black lines, all 
$t_\mathrm{coll}$ as orange ones. 
The increase of $t_\mathrm{coll}$ 
inside  $80$\,AU  
is due to depletion of 
dust in this region.}
\label{fig:tremov}
\end{figure} 

Nevertheless, the end result is the same as if the particles had been fully
eroded: they are all eventually removed from the system.
Moreover, for the bigger grains, the timescale $t_\mathrm{rem}$ for this removal
(which we conservatively define as the time it takes for a grain
at initial distance distance $R_0$ to migrate beyond $2R_0$) is
comparable to $t_\mathrm{sp}$ estimated in the previous section. Indeed,
for a grain
in the submillimetre to millimetre range, most of the sputtering erosion
takes place at a constant distance from the star.
It is then followed by a relatively brief ejection phase, which
begins once its size has dropped below the $\beta \approx 0.5$ limit, so
that the total time from start to ejection is almost equal to the erosion
time $t_{\mathrm{sp}_{1/2}}$ to reach $\beta=0.5$ 
($s_{1/2}=4\,\mu$m).
This time $t_{\mathrm{sp}_{1/2}}$ is in turn almost equal to the full erosion
time $t_\mathrm{sp}$ (the erosion time from $4\,\mu$m to 0 being negligible
compared to $t_{\mathrm{sp}_{1/2}}$). For grains with smaller initial $s_0$,
the situation is less simple, the time required to
reach $s_{1/2}=4\,\mu$m getting significantly shorter than $t_\mathrm{sp}$ when
$s_0$ is close to $s_{1/2}$. Furthermore, the ejection timescale
$t_\mathrm{ej}$ is no longer negligible compared to 
$t_{\mathrm{sp}_{1/2}}$
and has to be taken into account when estimating $t_\mathrm{rem}$.
For even smaller unbound grains, of course, sputtering plays only a very
minor role. These particles are ejected before any significant
erosion takes place, so that $t_\mathrm{rem}\simeq t_\mathrm{ej}$.

A set of additional runs has been carried out, where $t_\mathrm{rem}$ is estimated,
for the $1\,\mu$m to 5\,mm size range, as a function of initial distance from
the star (Fig.~\ref{fig:tremov}). As expected, differences with
the ``static'' case increase as $s_0$ gets smaller. 
For  the smallest grains, $t_\mathrm{rem}$ is much shorter than $t_\mathrm{sp}$
and is comparable to the orbital timescale at 100\,AU, i.e. 
$\sim 10^{3}$\,yr.

 \subsubsection{Comparison with collision timescales}

In Fig.~\ref{fig:tremov}, we present a summary of all our results
regarding ice survival times in the $\beta$~Pic system.
A first conclusion is that in the region where most of the
dust is observed ($80\la R \la 150$\,AU), sputtering is always
the dominant ice removal mechanism.
We also clearly see what has been already mentioned in the previous
subsections, i.e.\ that for all particles in the dust size
range, ice survival timescales $t_{\mathrm{ice}}$ (be it due to
sublimation or sputtering erosion combined to dynamical ejection)
are always shorter than the system's age. We confirm the result
derived in the static case, i.e.\ that
the smallest icy grains which can survive in the whole
$>80$\,AU region over $\simeq 10^{7}$\,yr have a size 
$s_{\mathrm{surv}}\simeq5$\,mm.

However, the system's age is probably not the only relevant
timescale, $t_{\mathrm{ice}}$ has to be compared to. Indeed, it
is well established that no debris disc grains are primordial,
but that they should be continuously resupplied by collisions
among bigger and undetectable objects 
\citep[e.g.,][ for a detailed discussion on the subject]{theb03}.
In this case, a small and (partially) icy grain could be
produced from a bigger particle for which
$t_{\mathrm{ice}}>t_{\mathrm{age}}$ and possibly be destroyed by another collision
before sputtering is able to fully  erode 
it.\footnote{We here consider ``erosion'' in the extended sense
of erosion to the $s_{1/2}$ limit followed by dynamical ejection.}

A first issue is here to see if collisions cannot by themselves
evaporate all ices from a grain's surface. 
From the estimates given in Table I of \citet{Tielens94},
the threshold impact velocity for achieving vaporization of water ice
is $v_{\mathrm{th}}\simeq 6.5$\,km\,s$^{-1}$. This value is
very high: it exceeds the orbital velocity beyond $\simeq25$\,AU, thus
making the occurrence of collisions at $v>v_{\mathrm{th}}$ very unlikely
(except possibly for very small unbound ($\beta > 0.5$ grains).

The rate at which dust grains are produced and destroyed
by collisions should in principle be estimated with
proper numerical simulations which exceed by far the scope of this paper
\citep[see e.g.][]{theb03,kriv06,theb07}.
We will here consider the collisional 
lifetime\footnote{The collisional lifetime being defined as the time
it takes for a given particle of size $s$ to lose 100\% of its initial mass, be
it by  accumulation of numerous partially eroding (cratering)
impacts, or by one violent fragmenting (i.e.\ total shattering) event.
If the system is in collisional equilibrium, $t_{\mathrm{coll}}$
is also equal to the production lifetime, i.e. the time it takes
to produce (by collisions among bigger bodies) the total number of  size $s$ 
particles present in the system.}
semi-empirical formula numerically
derived by \citet{theb07} for extended debris discs, valid
in the $ s \la 10^{6}s_{1/2}$ size range:

\begin{equation}
t_{\mathrm{coll}}= t_{\mathrm{coll0}}\,\left[\left(\frac{s}{s_1}\right)^{-2}
+\left(\frac{s}{s_2}\right)^{2.7}\right] \,\,\, {\rm for}\,\,\,  s<s_2
\label{tc1} 
\end{equation}
with $s_1=1.2s_{1/2}$ and $s_2=100s_{1/2}$, 
and
\begin{equation}
t_{\mathrm{coll}}= t_{\mathrm{coll0}}\,\left(\frac{s}{s_2}\right)^{0.3}
\,\,\,\,\,\,\,\,\, {\rm for}\,\,\,  s>s_2
\label{tc2} 
\end{equation}
where $t_{\mathrm{coll0}}=(\tau\Omega)^{-1}$ is the usual, and often misleading
reference collision timescale directly derived from
the vertical optical depth $\tau$ ($\Omega$ being the angular
orbital velocity). For the radial distribution of the optical
depth, we follow \citet{GTA07} and assume the best fit obtained by
\citet{Au01}.

As can be seen from Fig.~\ref{fig:tremov}, the situation is a complex one.
To better understand the role of collisions, let us
here make the simplifying but reasonable assumption that a particle
of a given size $s$ is mostly produced from collisions involving
parent bodies which are typically of size 
$s_\mathrm{parent}\simeq2-5\,s$ \citep{theb03}.
Let us now consider as a reliable starting point
the smallest particles that are able
to retain ices over the system's age regardless of the
additional effect of collisions, i.e. the ones with 
$t_{\mathrm{ice}}>t_{\mathrm{age}}$.
As already discussed, this corresponds to objects with size
$s_{\mathrm{surv}}\simeq 5$\,mm.
Fragments produced from these parent bodies, which are
predominantly in the $\simeq 1$\,mm range, should thus initially
contain ices. Since for these 1\,mm objects $t_{\mathrm{coll}}<t_{\mathrm{ice}}$
almost everywhere in the disc (beyond $\geq 40$\,AU), we can assume
that most of these grains will remain icy over their collisionally-imposed
lifetime. If we carry on with this iterative procedure, we see that
we can work our way down to $s\simeq 20\,\mu$m before $t_{\mathrm{coll}}$ begins
to get comparable to $t_{\mathrm{ice}}$ in some regions of the system.
For all sizes below this limit, ice survival times are smaller
than collisional lifetimes, 
implying it is unlikely that collisionally
produced fragments can remain icy. 
As a consequence,  $20\,\mu$m  may be considered as the approximate
lower boundary for grains which can have an icy component in the
$\beta$~Pic disc, at least in the 50--150\,AU region where most
of the dust has been detected.

Of course these results should be taken with great care, since
this very complex issue should in principle be addressed using
full scale collision evolution simulations. Such detailed
study will be the purpose of a forthcoming paper.

\subsubsection{Gas production rate}

For the specific case of $\beta$~Pic, these studies of
photosputtering and its consequences can be taken one step
further than estimating grain sizes which $could$ be icy. Indeed,
as mentioned in Sect.~\ref{sec:gas} sublimation and photosputtering 
do not simply ``remove'' ice from
the system, they transform it into gas.
Water molecules desorbed from the surface are 
quickly dissociated in gas phase by the stellar UV radiation
 into H and O \citep{Li98}.
Using the
assumptions of Sect.~\ref{sec:uvsputt} together with the 
dust distribution derived by \citet{Au01}, we estimate 
that, should most of the grains be icy ($\eta=1$), 
the corresponding production rate of gas  would be
$\dot{M}_{\mathrm{H_2O}} \sim 10^{-6}\,M_{\oplus}$\,yr$^{-1}$,
and the column density production of oxygen 
$\dot{N}_{\mathrm{O}} \sim 10^{14}$\,cm$^{-2}$\,yr$^{-1}$.
Since oxygen is expected to be
neutral, it is also dynamically inert in the disc \citep{Fernandez06}.
 This means that oxygen released is not easily removed,
and will only slowly redistribute itself in the disc. 
By comparing to 
the observed O\,I column density 
$N_{\mathrm{O\,I}} = (5.5 \pm 2.5)\times 10^{15}$\,cm$^{-2}$ 
\citep{Roberge06}, we find that all the observed 
oxygen would be produced from photosputtering on an extremely short 
timescale, $\la 100$\,yr. Together with the fact that dust grains
unexposed by radiation are continuously produced in debris from
collisions, this implies that the parent bodies must be very poor in water 
ice. Otherwise, the efficiency of the photosputtering process
would quickly refill the system with unrealistically large amounts
of gaseous oxygen.

\subsection{Other systems  \label{sec:other}}

\begin{table}
\caption{UV sputtering rates and grain survival times,
numerically estimated for a sample of 11 debris disc systems.
All values are given at a reference distance of 100\,AU from
the central star, which is a characteristic scale 
of the density maximum in debris discs. 
The yield per incident photon is taken to be 
$Y=10^{-3} \times Q^{\prime}_\mathrm{abs}$.
 Systems marked with $^*$  have their sublimation boundaries 
outside 100\,AU. Thus the values given in the table do not 
reflect the grain lifetimes at 100\,AU, however they
can be scaled to farther distances. } 
\label{tab:sdot_other}
\begin{tabular}{lllll}
\hline\hline
 Name &Type   & $\dot{s}_\mathrm{sp}$\,($\mu$m~yr$^{-1}$)&
 \multicolumn{2}{c}{$t$\,(yr)}\\
 &&& 20$\,\mu$m & 1\,mm\\
\hline
HD66591$^*$ &B4V &    7 & (2)$^*$ & (140)$^*$ \\
HD142165$^*$ &B5V &    0.9 & (20)$^*$ & ($10^{3}$)$^*$ \\
 Vega$^*$&A0V 	&   0.09 &(250)$^*$ & ($10^{4}$)$^*$ \\
 HR\,4796A&A0V 	&  0.04 & 600 & 2$\times 10^{4} $\\
 Fomalhaut&A3V    & 3$\times 10^{-3}$ & 7$\times 10^{3}$ & 3$\times 10^{5}$ \\
 \bp &A5V & 	  4.5$\times 10^{-4}$ & 5.5$\times 10^{4} $& 2.3$\times10^{6} $\\
 HD\,40136& F1V &   5$\times 10^{-5}$ &  5$\times 10^{5}$ &2$\times 10^{7}$\\
  HD\,1581& G0V	&   $10^{-6}$&  2$\times 10^{7}$&  $10^{9}$ \\
  HD\,20807& G2V&   $10^{-6}$&  2$\times 10^{7}$&  $ 10^{9}$ \\
  $\epsilon$~Eridani&K2V  & 2$\times 10^{-7}$&  $ 10^{8}$&  5$\times 10^{9}$\\
  AU\,Mic &M1Ve	 &4$\times 10^{-7}$&   5$\times10^{7}$&  2.5$\times 10^{9}$\\
\hline
\end{tabular}
\end{table}

Performing similarly extensive studies for all debris disc systems is out of the
scope of the present paper.  We will here present, for a selection of 11
representative systems, estimates of $\dot{s}_\mathrm{sp}$ at a reference
distance of 100\,AU, as derived from Eq.\ref{eq:sdot_sp}. Furthermore, for each
system we present the numerically derived values for $t_\mathrm{rem}$, for a
``big'' 1\,mm and  a ``small'' $20\,\mu$m grain. As can be seen from
Table~\ref{tab:sdot_other}, $\dot{s}_\mathrm{sp}$ varies by 7 orders of
magnitude  depending on the stellar type.\footnote{UV fluxes were obtained 
from the MAST archive
http://archive.stsci.edu/index.html.
The systems with debris discs were selected from the 
Debris Disc Database http://www.roe.ac.uk/ukatc/research/topics/dust/identification.html}.

A can be clearly seen, for all considered stars, the sputtering removal
timescale of a $20\,\mu$m grain is always shorter than the age of the system. The only
exception is the M star AU Mic, where the erosion timescale is $\sim 50$\,Myr,
i.e. at least 4 times the age of the system (12\,Myr). For bigger, 1\,mm grains, the
situation is more favorable, at least around the less luminous G, K and M stars,
where the erosion timescale of a 1\,mm grain exceeds 1\,Gyr. However, for F stars and
upwards, millimetre-sized icy particles cannot survive for more than 
1--2$\times 10^{7}$\,yr, which is less than the minimum age of a typical debris
disc.

The above numerical estimates are presented to give a first estimate of
photosputtering effects as a function of  stellar type and should be interpreted
carefully. They are given at a reference distance of 100\,AU, whereas  dust
might be observed significantly closer or further away from the star.  This
limitation is, however, not too crucial since,  as a first approximation, the  erosion
rates and timescales we computed  can be rescaled with distance following a quadratic 
dependence of irradiation on $R$.  
Another issue that has not been addressed
here is the complex interplay between sputtering and collisions. Since for most
of the presented systems, dust size and spatial distribution is not as well
constrained as for $\beta$~Pic, such an exploration would be somewhat
tentative. It should, however, be kept in mind that intense collisional activity
could affect grain survivability. Furthermore, the microscopic
structure of the evaporating grain will certainly undergo change in the course 
of radiative de-icing. The resultant sputtered grain may be much more porous
than in the initial phase or, on the contrary, develop a microcometary 
structure in which dust is mixed with ice in the large core, and with an
altered organic ``crust'' near the surface \citep[a possibility mentioned in][]{Art97}. 
How the mechanical properties of a grain evolve with 
time in that case and how the phenomenon of dust avalanches is enhanced 
by icy nature of grains are new and potentially important questions.                              

Our preliminary conclusion is that, with the possible exception of
 M stars, there is no system where photosputtering effect can be 
 fully neglected.  Of
course, each of these systems has to be more carefully investigated, 
and our sputtering erosion estimates utilized in more advanced global dust evolution
models.

\section{Summary and Conclusion \label{sec:summary}} 

We have reconsidered the issue of icy grain presence and survival in
debris discs by quantitatively exploring the role of UV sputtering
in addition to the well known sublimation mechanism. 
Taking the archetypal $\beta$~Pic debris disc as a reference 
system, we show that sputtering is able to significantly erode
icy grains far beyond the sublimation-imposed snow line.
Using a dynamical model coupling grain erosion and orbital evolution,
we find that, in the 50--150\,AU region where the bulk of $\beta$~Pic's
grain population is thought to reside, all
grains smaller than a threshold size $s_{\mathrm{ero}}\sim 5$\,mm
(basically corresponding to all particles detectable by observations)
cannot retain their icy component over the age of the system.
Taking into account possible collisional activity slightly improves
the situation. There is indeed a non negligible size range below $s_{\mathrm{ero}}$
for which collision timescales are shorter than sputtering
erosion times, so that steady collisional reprocessing starting at
grains just above $s_{\mathrm{ero}}$ (in the icy ``reservoir'' size range)
could maintain a significant amount of smaller icy grains. 
However, when checking this icy grains hypothesis against the
observational constraints on gas abundances, we find that
the rate at which O\,I should be steadily produced by photosputtering
would very quickly lead to column densities exceeding by several
orders of magnitudes observational upper limits.
This seems like a relatively robust argument for ruling out
the assumption that, for the $\beta$~Pic system,
the observed grain population is predominantly icy.

We have explored the amplitude of the sputtering mechanism for 
other debris discs systems, deriving sputtering rates and
survival timescales, for different grain sizes,
 as a function of spectral type
We have shown that for all stars other than M dwarfs, no icy grain
in the $20\,\mu$m range can survive for more than $10^{7}$\,yr, while
millimetre-sized ice grains can survive in G stars and below.
These derivations are of course first order approximations. The obtained
erosion rates have to be included in more sophisticated dust
evolution models, taking for instance into account the coupled effect
of collisions and sputtering. This will be the purpose of a forthcoming study.

\begin{acknowledgements}
We thank Helen Fraser for providing detailed comments on the 
manuscript, and the referee Aigen Li for the careful review.
\end{acknowledgements}

\bibliographystyle{aa}
\bibliography{bibliography}

\begin{thebibliography}{49}
\expandafter\ifx\csname natexlab\endcsname\relax\def\natexlab#1{#1}\fi

\bibitem[{{Andersson} {et~al.}(2006){Andersson}, {Al-Halabi}, {Kroes}, \& {van
  Dishoeck}}]{Andersson06}
{Andersson}, S., {Al-Halabi}, A., {Kroes}, G.-J., \& {van Dishoeck}, E.~F.
  2006, ArXiv Astrophysics e-prints, 124

\bibitem[{{Artymowicz}(1988)}]{Art88}
{Artymowicz}, P. 1988, \apjl, 335, L79

\bibitem[{{Artymowicz}(1994)}]{Art94}
{Artymowicz}, P. 1994, in Circumstellar Dust Disks and Planet Formation, ed.
  R.~{Ferlet} \& A.~{Vidal-Madjar}, 47--+

\bibitem[{{Artymowicz}(1996)}]{Art96}
{Artymowicz}, P. 1996, in The Role of Dust in the Formation of Stars,
  Proceedings of the ESO Workshop Held at Garching, Germany, 11 - 14 September
  1995. Edited by Hans U. K{\"a}ufl and Ralf Siebenmorgen. Springer-Verlag
  Berlin Heidelberg New York. Also ESO Astrophysics Symposia (European Southern
  Observatory), p.137, ed. H.~U. {K{\"a}ufl} \& R.~{Siebenmorgen}, 137--+

\bibitem[{{Artymowicz}(1997)}]{Art97}
---. 1997, Annual Review of Earth and Planetary Sciences, 25, 175

\bibitem[{{Augereau} {et~al.}(2001){Augereau}, {Nelson}, {Lagrange},
  {Papaloizou}, \& {Mouillet}}]{Au01}
{Augereau}, J.~C., {Nelson}, R.~P., {Lagrange}, A.~M., {Papaloizou}, J.~C.~B.,
  \& {Mouillet}, D. 2001, \aap, 370, 447

\bibitem[{{Brandeker} {et~al.}(2004){Brandeker}, {Liseau}, {Olofsson}, \&
  {Fridlund}}]{Brandeker04}
{Brandeker}, A., {Liseau}, R., {Olofsson}, G., \& {Fridlund}, M. 2004, \aap,
  413, 681

\bibitem[{{Buck}(1981)}]{Buck96}
{Buck}, A.~L. 1981, Journal of Applied Meteorology, 20, 1527

\bibitem[{{Burns} {et~al.}(1979){Burns}, {Lamy}, \& {Soter}}]{BurnsLamy79}
{Burns}, J.~A., {Lamy}, P.~L., \& {Soter}, S. 1979, Icarus, 40, 1

\bibitem[{{Carlson}(1980)}]{Carlson80}
{Carlson}, R.~W. 1980, \nat, 283, 461

\bibitem[{{Chen} {et~al.}(2007){Chen}, {Li}, {Bohac}, {Kim}, {Watson}, {van
  Cleve}, {Houck}, {Stapelfeldt}, {Werner}, {Rieke}, {Su}, {Marengo},
  {Backman}, {Beichman}, \& {Fazio}}]{Chen07}
{Chen}, C.~H., {Li}, A., {Bohac}, C., {et~al.} 2007, \apj, 666, 466

\bibitem[{{Chen} {et~al.}(2006){Chen}, {Sargent}, {Bohac}, {Kim},
  {Leibensperger}, {Jura}, {Najita}, {Forrest}, {Watson}, {Sloan}, \&
  {Keller}}]{Chen06}
{Chen}, C.~H., {Sargent}, B.~A., {Bohac}, C., {et~al.} 2006, \apjs, 166, 351

\bibitem[{{Creech-Eakman} {et~al.}(2002){Creech-Eakman}, {Chiang}, {Joung},
  {Blake}, \& {van Dishoeck}}]{Creech02}
{Creech-Eakman}, M.~J., {Chiang}, E.~I., {Joung}, R.~M.~K., {Blake}, G.~A., \&
  {van Dishoeck}, E.~F. 2002, \aap, 385, 546

\bibitem[{{Di Folco} {et~al.}(2004){Di Folco}, {Th{\'e}venin}, {Kervella},
  {Domiciano de Souza}, {Coud{\'e} du Foresto}, {S{\'e}gransan}, \&
  {Morel}}]{dif04}
{Di Folco}, E., {Th{\'e}venin}, F., {Kervella}, P., {et~al.} 2004, \aap, 426,
  601

\bibitem[{{Dijkstra} {et~al.}(2003){Dijkstra}, {Dominik}, {Hoogzaad}, {de
  Koter}, \& {Min}}]{Dijkstra03}
{Dijkstra}, C., {Dominik}, C., {Hoogzaad}, S.~N., {de Koter}, A., \& {Min}, M.
  2003, \aap, 401, 599

\bibitem[{{Dominik} {et~al.}(2005){Dominik}, {Ceccarelli}, {Hollenbach}, \&
  {Kaufman}}]{Dominik05}
{Dominik}, C., {Ceccarelli}, C., {Hollenbach}, D., \& {Kaufman}, M. 2005,
  \apjl, 635, L85

\bibitem[{{Fanale} \& {Salvail}(1984)}]{Fanale84}
{Fanale}, F.~P. \& {Salvail}, J.~R. 1984, Icarus, 60, 476

\bibitem[{{Fern{\'a}ndez} {et~al.}(2006){Fern{\'a}ndez}, {Brandeker}, \&
  {Wu}}]{Fernandez06}
{Fern{\'a}ndez}, R., {Brandeker}, A., \& {Wu}, Y. 2006, \apj, 643, 509

\bibitem[{{Freudling} {et~al.}(1995){Freudling}, {Lagrange}, {Vidal-Madjar},
  {Ferlet}, \& {Forveille}}]{Freudling95}
{Freudling}, W., {Lagrange}, A.-M., {Vidal-Madjar}, A., {Ferlet}, R., \&
  {Forveille}, T. 1995, \aap, 301, 231

\bibitem[{{Galland} {et~al.}(2006){Galland}, {Lagrange}, {Udry}, {Chelli},
  {Pepe}, {Beuzit}, \& {Mayor}}]{Galland06}
{Galland}, F., {Lagrange}, A.-M., {Udry}, S., {et~al.} 2006, \aap, 447, 355

\bibitem[{{Gibb} {et~al.}(2004){Gibb}, {Whittet}, {Boogert}, \&
  {Tielens}}]{Gibb04}
{Gibb}, E.~L., {Whittet}, D.~C.~B., {Boogert}, A.~C.~A., \& {Tielens},
  A.~G.~G.~M. 2004, \apjs, 151, 35

\bibitem[{{Golimowski} {et~al.}(2006){Golimowski}, {Ardila}, {Krist},
  {Clampin}, {Ford}, {Illingworth}, {Bartko}, {Ben{\'{\i}}tez}, {Blakeslee},
  {Bouwens}, {Bradley}, {Broadhurst}, {Brown}, {Burrows}, {Cheng}, {Cross},
  {Demarco}, {Feldman}, {Franx}, {Goto}, {Gronwall}, {Hartig}, {Holden},
  {Homeier}, {Infante}, {Jee}, {Kimble}, {Lesser}, {Martel}, {Mei},
  {Menanteau}, {Meurer}, {Miley}, {Motta}, {Postman}, {Rosati}, {Sirianni},
  {Sparks}, {Tran}, {Tsvetanov}, {White}, {Zheng}, \& {Zirm}}]{Golimowski06}
{Golimowski}, D.~A., {Ardila}, D.~R., {Krist}, J.~E., {et~al.} 2006, \aj, 131,
  3109

\bibitem[{{Grigorieva} {et~al.}(2007){Grigorieva}, {Artymowicz}, \&
  {Th{\'e}bault}}]{GTA07}
{Grigorieva}, A., {Artymowicz}, P., \& {Th{\'e}bault}, P. 2007, \aap, 461, 537

\bibitem[{{Harrison}(1967)}]{Harrison67}
{Harrison}, e.~a. 1967, Science, 157

\bibitem[{{Jura} {et~al.}(1998){Jura}, {Malkan}, {White}, {Telesco}, {Pina}, \&
  {Fisher}}]{Jura98}
{Jura}, M., {Malkan}, M., {White}, R., {et~al.} 1998, \apj, 505, 897

\bibitem[{{Karmann} {et~al.}(2003){Karmann}, {Beust}, \& {Klinger}}]{Karmann03}
{Karmann}, C., {Beust}, H., \& {Klinger}, J. 2003, \aap, 409, 347

\bibitem[{{Krivov} {et~al.}(2006){Krivov}, {L{\"o}hne}, \& {Srem{\v
  c}evi{\'c}}}]{kriv06}
{Krivov}, A.~V., {L{\"o}hne}, T., \& {Srem{\v c}evi{\'c}}, M. 2006, \aap, 455,
  509

\bibitem[{{Lamy}(1974)}]{Lamy74}
{Lamy}, P.~L. 1974, \aap, 35, 197

\bibitem[{{Lecavelier des Etangs} {et~al.}(2001){Lecavelier des Etangs},
  {Vidal-Madjar}, {Roberge}, {Feldman}, {Deleuil}, {Andr{\'e}}, {Blair},
  {Bouret}, {D{\'e}sert}, {Ferlet}, {Friedman}, {H{\'e}brard}, {Lemoine}, \&
  {Moos}}]{Lecavelier01}
{Lecavelier des Etangs}, A., {Vidal-Madjar}, A., {Roberge}, A., {et~al.} 2001,
  \nat, 412, 706

\bibitem[{{Li} \& {Greenberg}(1998{\natexlab{a}})}]{Li98}
{Li}, A. \& {Greenberg}, J.~M. 1998{\natexlab{a}}, \aap, 331, 291

\bibitem[{{Li} \& {Greenberg}(1998{\natexlab{b}})}]{Li98b}
---. 1998{\natexlab{b}}, \aap, 338, 364

\bibitem[{{Li} {et~al.}(2003){Li}, {Lunine}, \& {Bendo}}]{Li03Eri}
{Li}, A., {Lunine}, J.~I., \& {Bendo}, G.~J. 2003, \apjl, 598, L51

\bibitem[{{Lien}(1990)}]{Lien90}
{Lien}, D.~J. 1990, \apj, 355, 680

\bibitem[{{Lisse} {et~al.}(2006){Lisse}, {Beichman}, {Bryden}, \&
  {Wyatt}}]{Lisse07}
{Lisse}, C.~M., {Beichman}, C.~A., {Bryden}, G., \& {Wyatt}, M.~C. 2006, ArXiv
  Astrophysics e-prints

\bibitem[{{Marti} \& {Mauersberger}(1993)}]{Marti93}
{Marti}, J. \& {Mauersberger}, K. 1993, \grl, 20, 363

\bibitem[{{Mauersberger} \& {Krankowsky}(2003)}]{Mauersberger03}
{Mauersberger}, K. \& {Krankowsky}, D. 2003, \grl, 30, 21

\bibitem[{{Meeus} {et~al.}(2001){Meeus}, {Waters}, {Bouwman}, {van den Ancker},
  {Waelkens}, \& {Malfait}}]{Meeus01}
{Meeus}, G., {Waters}, L.~B.~F.~M., {Bouwman}, J., {et~al.} 2001, \aap, 365,
  476

\bibitem[{{Roberge} {et~al.}(2006){Roberge}, {Lecavelier Des Etangs},
  {Vidal-Madjar}, {Feldman}, {Deleuil}, {Bouret}, \& {Ferlet}}]{Roberge06}
{Roberge}, A., {Lecavelier Des Etangs}, A., {Vidal-Madjar}, A., {et~al.} 2006,
  in ASP Conf. Ser. 348: Astrophysics in the Far Ultraviolet: Five Years of
  Discovery with FUSE, ed. G.~{Sonneborn}, H.~W. {Moos}, \& B.-G. {Andersson},
  294--+

\bibitem[{{Ruscic} {et~al.}(2002){Ruscic}, {Wagner}, {Harding}, {Asher},
  {Feller}, {Dixon}, {Peterson}, {Song}, {Qian}, {Ng}, {Liu}, {Chen}, \&
  {Schwenke}}]{Ruscic02}
{Ruscic}, B., {Wagner}, A.~F., {Harding}, L.~B., {et~al.} 2002, J. Phys. Chem.
  A., 106, 2727

\bibitem[{{Smith} \& {Terrile}(1984)}]{Smith84}
{Smith}, B.~A. \& {Terrile}, R.~J. 1984, Science, 226, 1421

\bibitem[{{Smithsonian Meterological Tables}(1984)}]{Smithonian84}
{Smithsonian Meterological Tables}. 1984, in 5th ed., 350

\bibitem[{{Tamura} {et~al.}(2006){Tamura}, {Fukagawa}, {Kimura}, {Yamamoto},
  {Suto}, \& {Abe}}]{Tamura06}
{Tamura}, M., {Fukagawa}, M., {Kimura}, H., {et~al.} 2006, \apj, 641, 1172

\bibitem[{{Telesco} {et~al.}(2005){Telesco}, {Fisher}, {Wyatt}, {Dermott},
  {Kehoe}, {Novotny}, {Mari{\~n}as}, {Radomski}, {Packham}, {De Buizer}, \&
  {Hayward}}]{Telesco05}
{Telesco}, C.~M., {Fisher}, R.~S., {Wyatt}, M.~C., {et~al.} 2005, \nat, 433,
  133

\bibitem[{{Th{\'e}bault} \& {Augereau}(2005)}]{Thebault05}
{Th{\'e}bault}, P. \& {Augereau}, J.-C. 2005, \aap, 437, 141

\bibitem[{{Th{\'e}bault} \& {Augereau}(2007)}]{theb07}
---. 2007, \aap, 472, 169

\bibitem[{{Th{\'e}bault} {et~al.}(2003){Th{\'e}bault}, {Augereau}, \&
  {Beust}}]{theb03}
{Th{\'e}bault}, P., {Augereau}, J.~C., \& {Beust}, H. 2003, \aap, 408, 775

\bibitem[{{Tielens} {et~al.}(1994){Tielens}, {McKee}, {Seab}, \&
  {Hollenbach}}]{Tielens94}
{Tielens}, A.~G.~G.~M., {McKee}, C.~F., {Seab}, C.~G., \& {Hollenbach}, D.~J.
  1994, \apj, 431, 321

\bibitem[{{Westley} {et~al.}(1995){Westley}, {Baragiola}, {Johnson}, \&
  {Baratta}}]{Westley95}
{Westley}, M.~S., {Baragiola}, R.~A., {Johnson}, R.~E., \& {Baratta}, G.~A.
  1995, \planss, 43, 1311

\bibitem[{{Zuckerman} {et~al.}(2001){Zuckerman}, {Song}, {Bessell}, \&
  {Webb}}]{zuc01}
{Zuckerman}, B., {Song}, I., {Bessell}, M.~S., \& {Webb}, R.~A. 2001, \apjl,
  562, L87

\end{thebibliography}

\end{document}